\documentclass[onecolumn,doublespace,12pt fonts,times]{aastex6}

\newcommand{\ag}{     {Ann. Geophys.}}
\newcommand{\jastp}{  {J. Atmos. Solar-Terr. Phys.}}

\shorttitle{Collision nature of the CMEs}
\shortauthors{Mishra et al.}

\usepackage{amsmath}
\usepackage{graphicx}
\usepackage{float}

\begin{document}

\title{On Understanding the Nature of Collision  of Coronal Mass Ejections Observed by \textit{STEREO}}

\author{Wageesh Mishra\altaffilmark{1}, Yuming Wang\altaffilmark{1} and Nandita Srivastava\altaffilmark{2,}\footnote{Centre for Excellence in Space Sciences, India, \url{http://www.cessi.in}}}


\altaffiltext{1}{CAS Key Laboratory of Geospace Environment, Department of Geophysics and Planetary Sciences,\\
University of Science and Technology of China, Hefei 230026, China; wageesh@ustc.edu.cn, ymwang@ustc.edu.cn}
\altaffiltext{2}{Udaipur Solar Observatory, Physical Research Laboratory, Badi Road, Udaipur 313001, India}

\begin{abstract}
Our study attempts to understand the collision characteristics of two coronal mass ejections (CMEs) launched successively from the Sun  on 2013 October 25.  The estimated kinematics, from three-dimensional (3D) reconstruction techniques applied to observations of CMEs by SECCHI/Coronagraphic (COR) and Heliospheric Imagers (HIs), reveal their collision around 37 $R_\sun$ from the Sun.  In the analysis, we take into account the propagation and expansion speeds, impact direction, angular size as well as the masses of the CMEs.  These parameters are derived from imaging observations, but may suffer from large uncertainties. Therefore, by adopting head-on as well as oblique collision scenarios, we have quantified the range of uncertainties involved in the calculation of the coefficient of restitution for expanding magnetized plasmoids. Our study shows that the comparatively large expansion speed of the following CME than that of the preceding CME, results in a higher probability of super-elastic collision. We also infer that a relative approaching speed of the CMEs lower than the sum of their expansion speeds increases the chance of super-elastic collision. The analysis under a reasonable errors in observed parameters of the CME, reveals the larger probability of occurrence of an inelastic collision for the selected CMEs. We suggest that the collision nature of two CMEs should be discussed in 3D, and the calculated value of the coefficient of restitution may suffer from a large uncertainty. 
\end{abstract}


\section{Introduction} \label{sec:intro}
Coronal mass ejections (CMEs) being the most energetic events on the Sun are expanding magnetized plasma blobs in the heliosphere. If they reach the Earth with a southward directed magnetic field orientation, they can cause intense geomagnetic storms \citep{Dungey1961,Gosling1993,Gonzalez1994}. They are frequently launched from the Sun, especially during solar maximum when their interaction or collision in the heliosphere is possible. Historically, such interaction was inferred using in situ data from \textit{Pioneer 9} and twin \textit{Helios} spacecraft \citep{Intriligator1976,Burlaga1987}.  However, the first observational evidence was provided by \citet{Gopalswamy2001apj} using Large Angle and Spectrometric COronagraph (LASCO; \citealp{Brueckner1995}) on-board \textit{SOlar and Heliospheric Observatory (SOHO)} and long wavelength radio observations. It has been suggested that some interacting CMEs have long interval of strong southward magnetic field and can produce major disturbances in the Earth's magnetosphere \citep{Wang2003a,Farrugia2004,Farrugia2006,Lugaz2014}. 

Before the \textit{Solar TErrestrial RElations Observatory} (\textit{STEREO}) \citep{Kaiser2008} era, CMEs could only be imaged near the Sun from one viewpoint of SOHO and we lacked their 
3D kinematics. Therefore, understanding CME-CME interaction was mainly based on magnetohydrodynamics (MHD) numerical simulations studies \citep{Vandas1997,Vandas2004,Gonzalez-Esparza2004,Lugaz2005,Wang2005,Xiong2006,Xiong2007,Xiong2009}. With the availability of wide angle imaging observations of Heliospheric Imagers (HIs) on-board \textit{STEREO} from multiple viewpoints, several cases of interacting CMEs have been recently reported in the literature \citep{Harrison2012,Liu2012,Lugaz2012,Mostl2012,Martinez-Oliveros2012,Shen2012,Temmer2012,Webb2013,Mishra2014a,Mishra2015,Colaninno2015}.  Also, the simulations based studies on the observed cases of CMEs are also being done to advance our understanding of such interaction \citep{Lugaz2013,Shen2013,Shen2014,Niembro2015,Shen2016}.

Understanding the interaction of CMEs is of interest because of their impact on many areas of heliospheric research.  Several cases of CME-CME interaction studies have focused on understanding their nature of collision, particle acceleration and geoeffectiveness \citep{Shen2012,Lugaz2014,Ding2014}.  Interacting CMEs also provide an unique opportunity to study the evolution of the shock strength, structure and its effect on plasma parameters of preceding CME \citep{Wang2003a,Lugaz2005,Mostl2012,Liu2012,Lugaz2015}. It is suggested that due to preconditioning of ambient medium by preceding CME, any following CME may experience high \citep{Temmer2012,Mishra2015a} or low drag \citep{Temmer2015} before their noticeable collision or merging. We use the term ``interaction'' and ``collision'' for two different sense as defined in \citep{Mishra2014a}.  By ``interaction'' we mean a probable exchange of momentum between the CMEs is in progress, however we could not notice obvious joining of their features in the imaging observations. The ``collision'' stands for the scenario noticed in imaging observation, where two CMEs moving with different speeds come in close contact with each other and show an exchange of momentum till they achieve an approximately equal speed or they get separated from each other.  Colliding CMEs can display change in their kinematics and morphology after the collision, and hence the prediction of their arrival time to Earth becomes challenging.  The knowledge about nature of collision of CMEs may be utilized to predict their post-collision kinematics.

Using twin viewpoint \textit{STEREO} observations, more accurate estimation of kinematics and masses of CMEs is possible, however, recent case studies are not in agreement about the nature of collision of the CMEs. This disagreement is possible as each case study have taken different candidate CMEs having probably different characteristics. Some studies exploiting imaging observations have shown a super-elastic collision of CMEs \citep{Shen2012,Colaninno2015} while some advocate inelastic \citep{Mishra2015} or close to elastic collision \citep{Mishra2014a,Mishra2015a}. This poses a question as to what determines the nature of collision, i.e. coefficient of restitution to vary from super-elastic to inelastic range. Most of the earlier studies have considered a simplistic approach that CMEs are propagating exactly in the same direction (i.e. head-on collision), and also have not taken the expansion speed or angular size of CME into account \citep{Mishra2014a,Mishra2015,Mishra2015a}. \citet{Schmidt2004} have studied the obliquely colliding CMEs using numerical  simulation. Some earlier studies have suggested the collision nature of the CMEs based on their deflection and change in the dynamics without explicit mentioning the value of coefficient of restitution \citep{Lugaz2012,Temmer2012,Colaninno2015}.  \citet{Shen2012} for the first time studied the oblique collision of CMEs using imaging observation, and considered several uncertainties into account, however they did not discuss on constraining the conservation of momentum. The straightforward use of observed CME characteristics (speed and mass) which may involve large errors, may be a reason for conservation of momentum to be no longer valid. We admit that previous studies are away from the real scenario and each has different limitations.

Taking an exception to \citet{Shen2012}, we are not aware of other study which thoroughly discusses the uncertainties involved in understanding the nature of collision of the CMEs.  Hence, we take next step to address the limitations of previous studies, and to investigate the role of characteristics of CMEs, e.g. direction, mass, propagation speed, expansion speed and angular size, on the collision nature. For this purpose, we selected two CMEs which occurred almost 7 hr apart on 2013 October 25 and collided with each other in HI-1 field-of-view. The collision around such a moderate distance from the Sun is well suited to our collision picture. This is because near the Sun coronal magnetic structures may interfere the CME dynamics and accurate estimation of the dynamics far from the Sun using HI observations is difficult \citep{Howard2011,Davies2012,Davies2013,Mishra2014}. We apply the Graduated Cylindrical Shell (GCS) fitting technique \citep{Thernisien2009} on coronagraphic images and Self-Similar Expansion (SSE) method \citep{Davies2012} on HI images of the CMEs to estimate their kinematics. This is discussed in Section~\ref{recon} including a description on estimation of true mass of the CMEs using \citet{Colaninno2009} method and identification of collision phase from the kinematics profile. Section~\ref{coli} presents the analysis and results from the head-on and oblique collision scenario, and shows the limitations of the approach of simplistic head-on collision undertaken in earlier studies. The various limitations of the present study are discussed in Section~\ref{Dis} and conclusions are presented in Section~\ref{Res}.

\section{Tracking of CMEs in the heliosphere} \label{recon}
Tracking of CME from its lift-off in the corona to the Earth or even beyond is possible using the imaging instruments of \textit{Sun Earth Connection Coronal and Heliospheric Investigation} (SECCHI) package on-board \textit{STEREO} spacecraft.   In the following section, we track the heliospheric evolution of CMEs from different viewpoints and apply suitable 3D reconstruction techniques to estimate their kinematics. 

\begin{figure}
\begin{center}
\includegraphics[angle=0,scale=0.45]{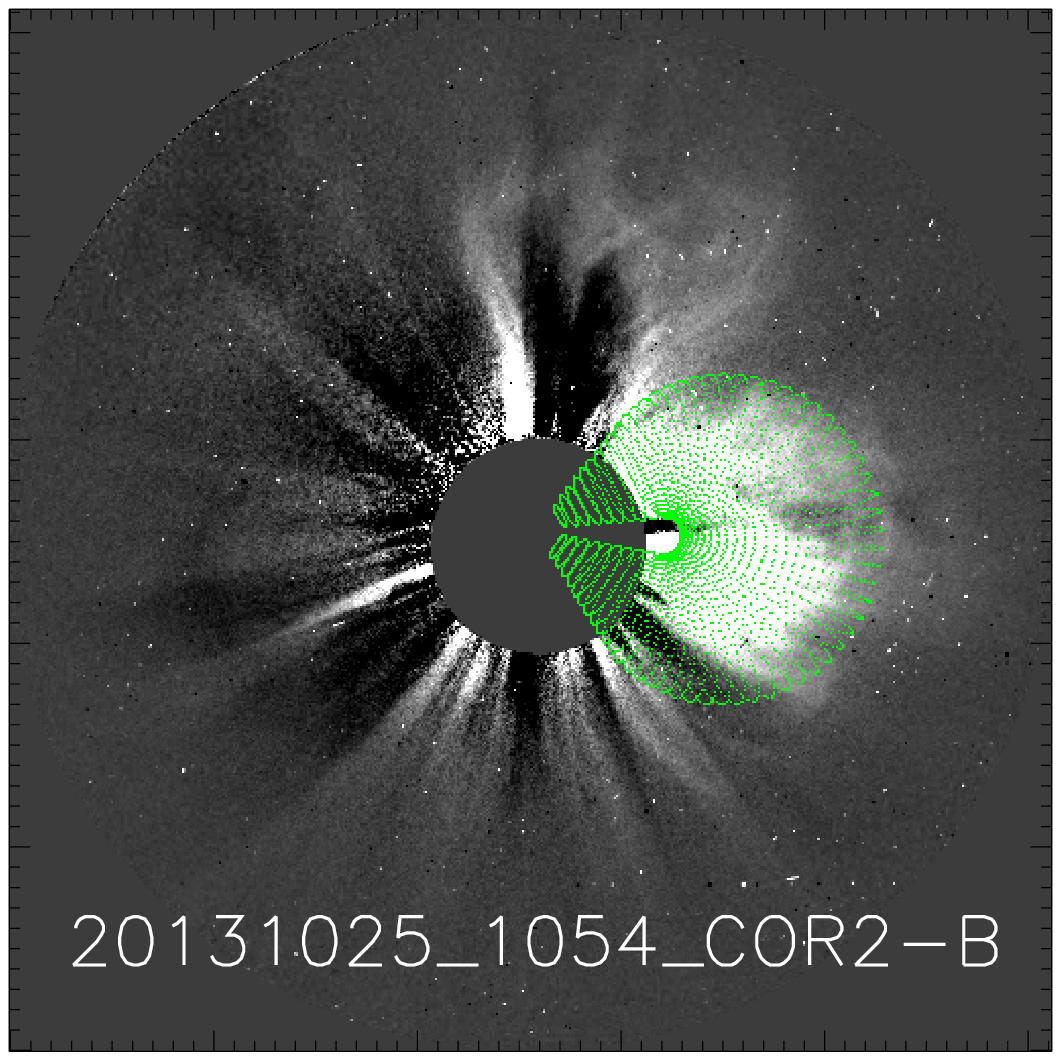}
\includegraphics[angle=0,scale=0.45]{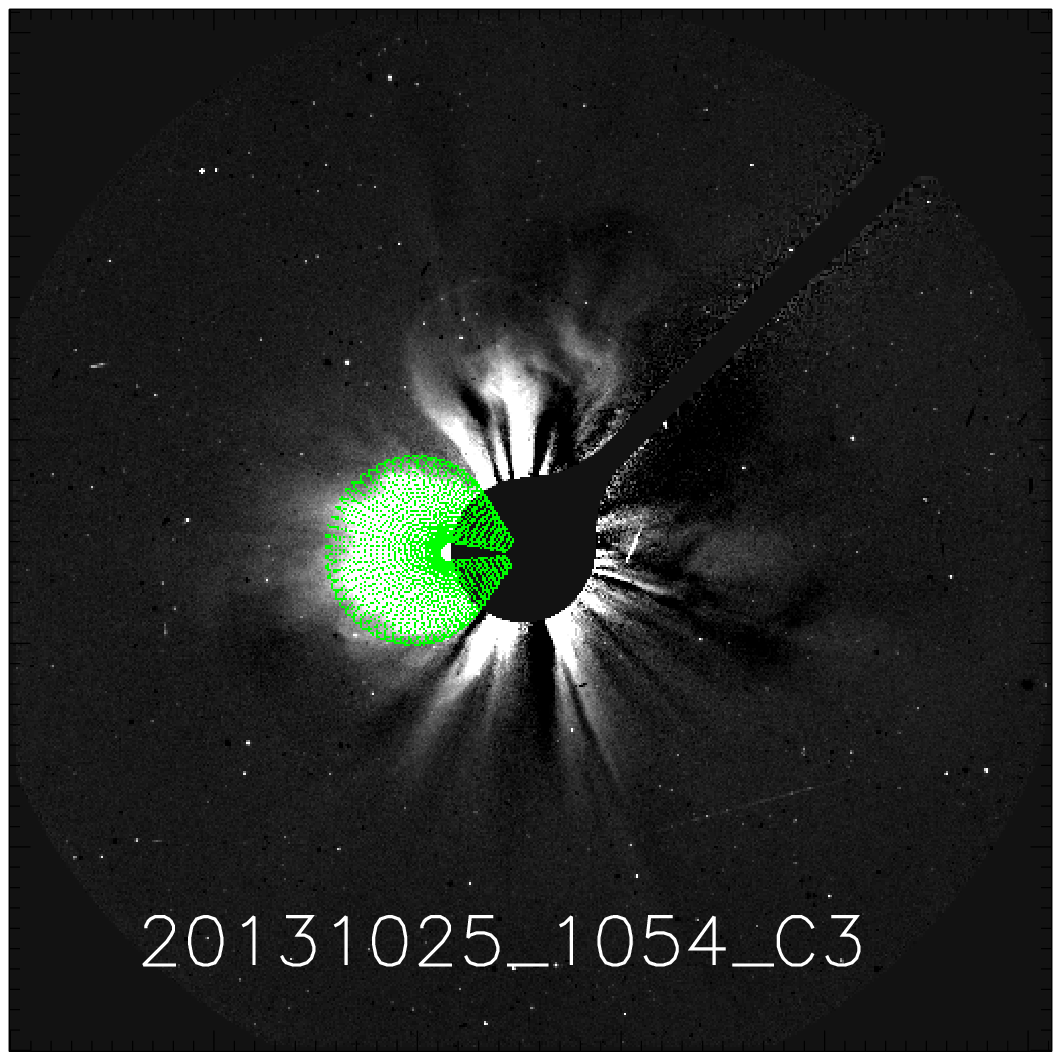}
\includegraphics[angle=0,scale=0.45]{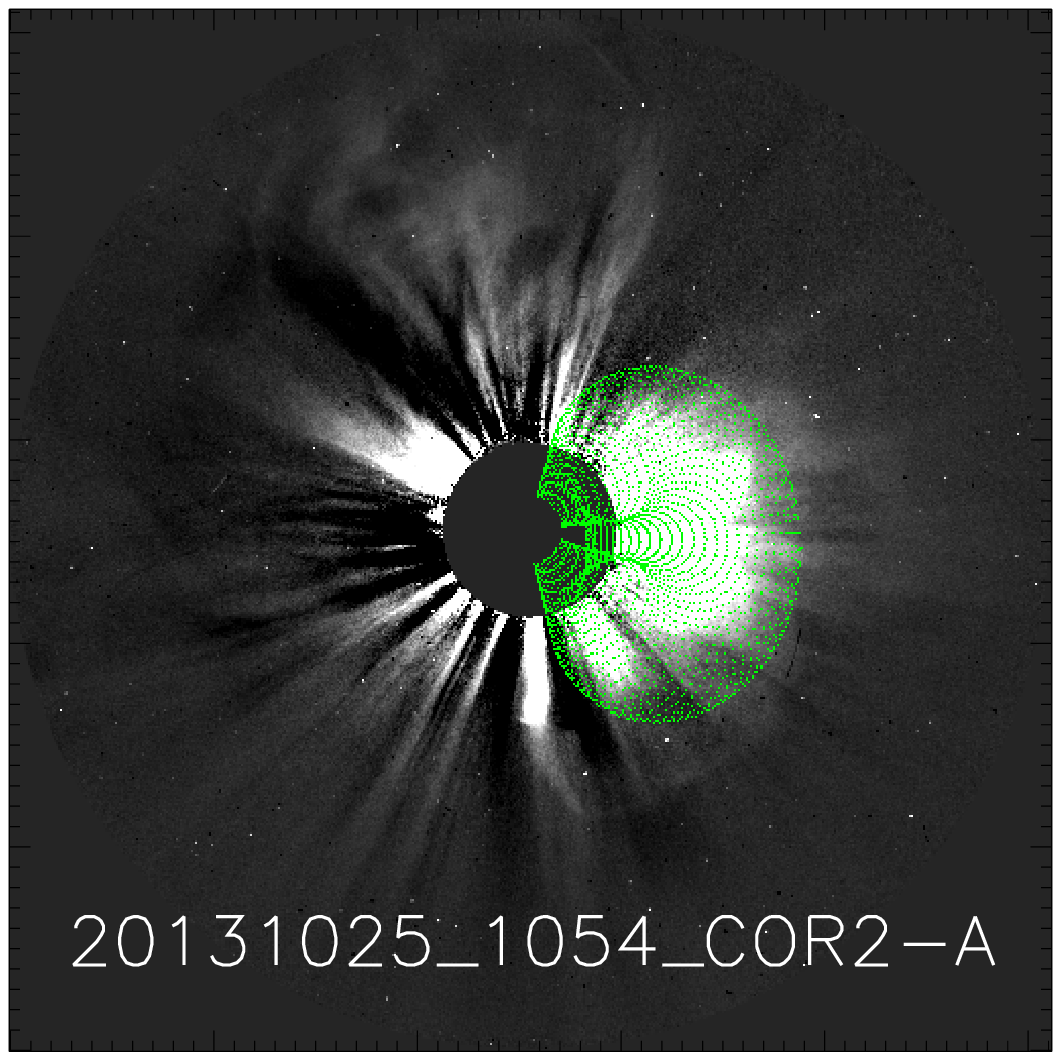}\\
\includegraphics[angle=0,scale=0.45]{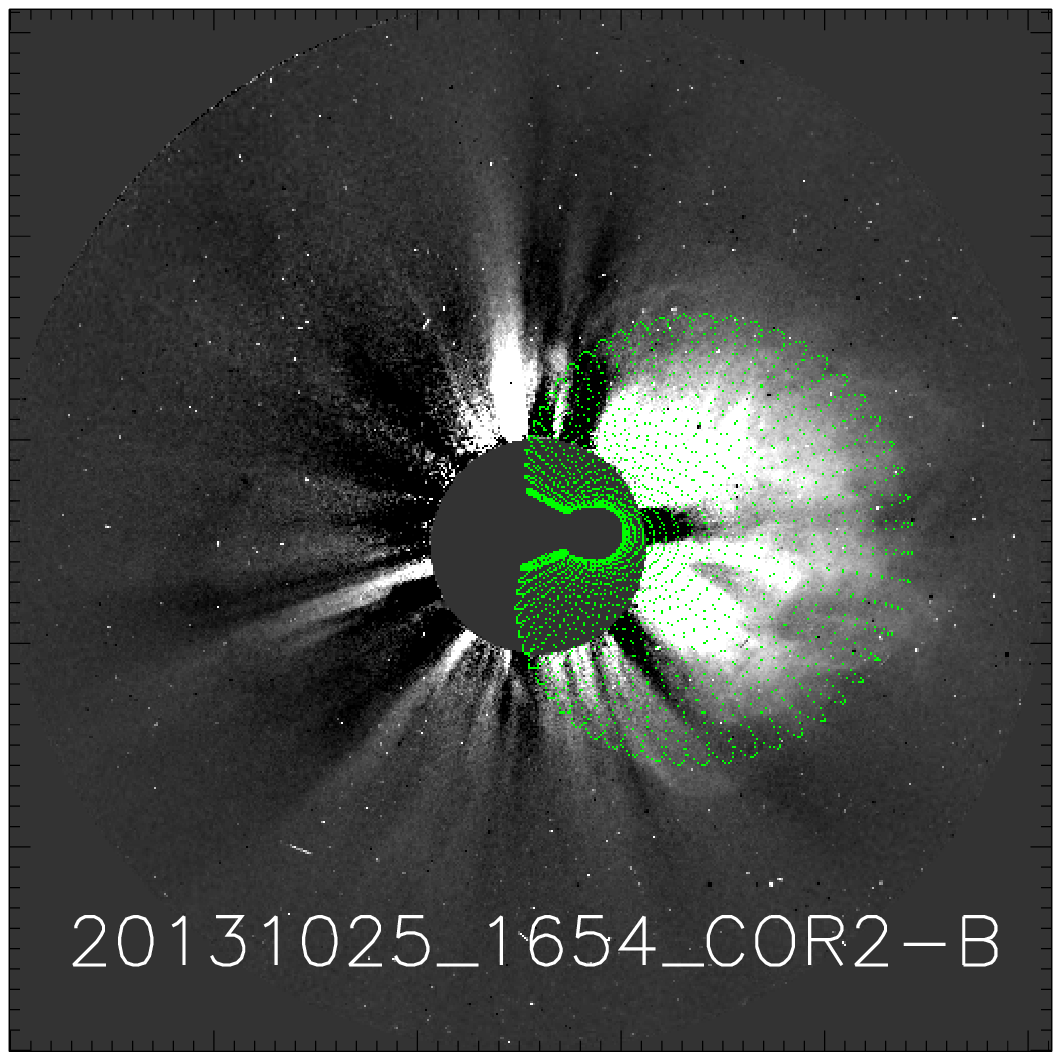}
\includegraphics[angle=0,scale=0.45]{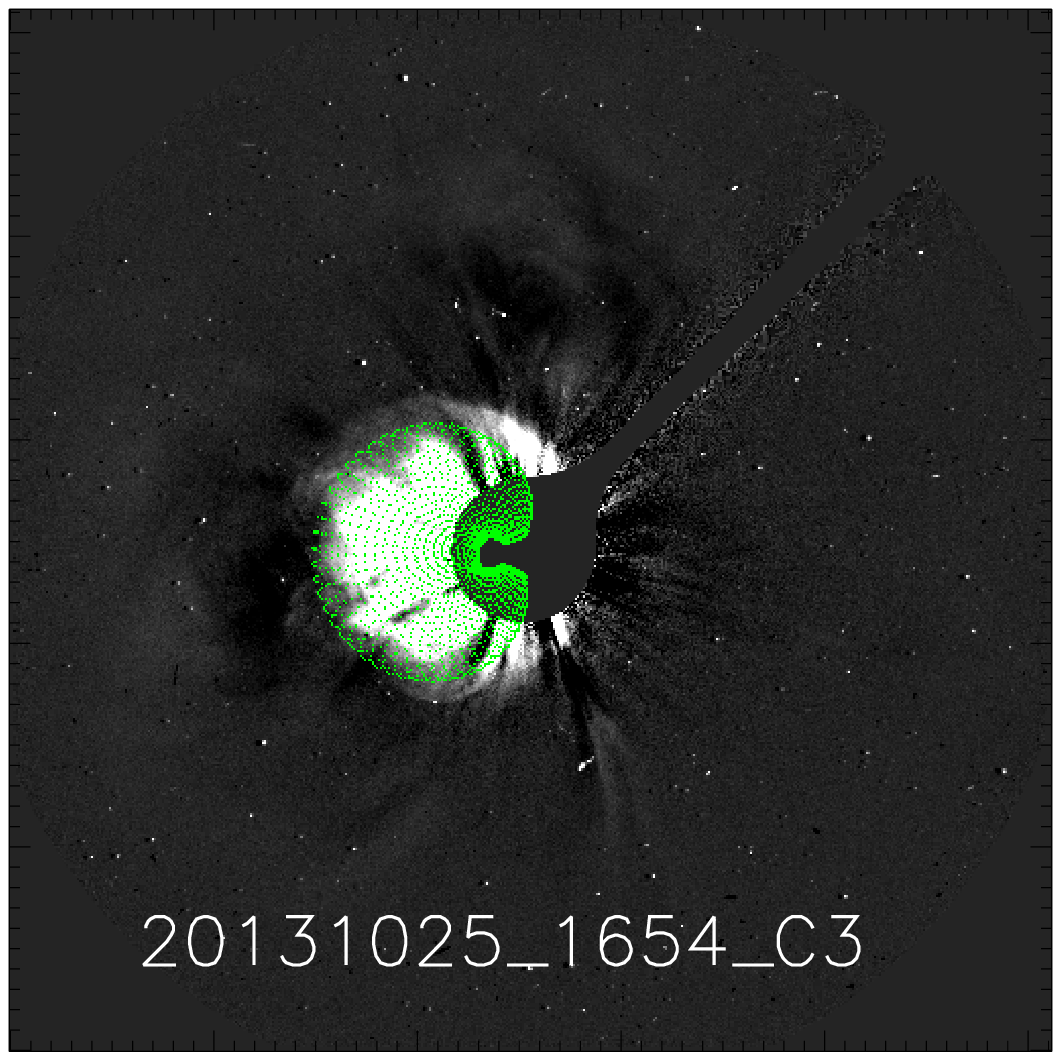}
\includegraphics[angle=0,scale=0.45]{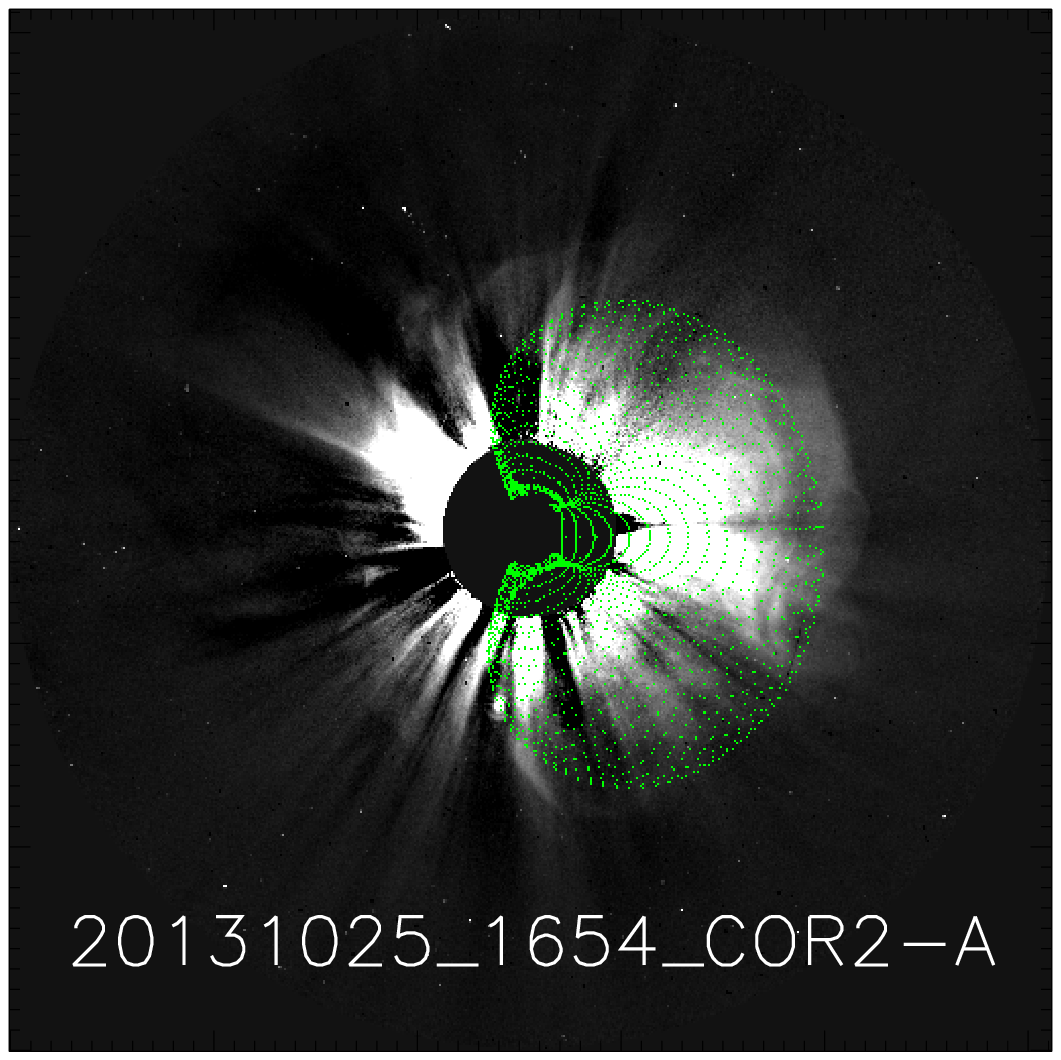}\\
\caption{The GCS model wire-frame with green is overlaid on the images of CME1 (top panels) and CME2 (bottom panels). The triplet of concurrent images are taken from \textit{STEREO}/COR2-B (left), SOHO/LASCO-C3 (middle), and \textit{STEREO}/COR2-A (right) around 10:54 UT for CME1 and  16:54 UT on October 25 for CME2, respectively.}
\label{GCS}
\end{center}
\end{figure}

\subsection{Estimation of kinematics in COR2 field-of-view}
\label{reconcor}
The selected CMEs in our study were recorded as halo CMEs from \textit{SOHO}/LASCO-C2 around 8:15 UT and 15:15 UT on 2013 October 25, respectively. We term these subsequently launched preceding and following CMEs as CME1 and CME2, respectively. To estimate the 3D kinematics of CMEs, we have applied the GCS  forward fitting model \citep{Thernisien2009} to the contemporaneous  images of the CMEs obtained from the SECCHI/COR2-B, SOHO/LASCO-C3 and SECCHI/COR2-A coronagraphs. Figure~\ref{GCS} shows images of  CME1 and CME2 overlaid with the fitted GCS wire-framed contour (in green). From this method, we note the propagation direction of CME1 along E70N03 (within error of $\pm$ 5$^\circ$) at a distance of 11.5$\pm$1.0 R$_\odot$. The propagation direction for the following CME2 is along the E65N03 (within an error of $\pm$ 5$^\circ$) at a distance of 12.5$\pm$1.0 R$_\odot$. In addition to aforementioned propagation direction, the best visual GCS fitting gives a half angle of 30$\pm$5$^{\circ}$, a tilt angle of 90$\pm$20$^{\circ}$, and an aspect ratio of 0.39$\pm$0.10 for CME1. The half angle, tilt angle and aspect ratio for CME2 is 65$\pm$5$^{\circ}$, 90$\pm$20$^{\circ}$, 0.59$\pm$0.10, respectively. The 3D speed of 
CME1 is noted as 485 km s$^{-1}$ and for CME2 it is 1000 km s$^{-1}$. The longitudes of CME1 (i.e. $\phi_{1}$=-70$^{\circ}$) and CME2 (i.e. $\phi_{2}$=-65$^{\circ}$) and their speeds suggest their propagation eastward from the Sun-Earth line, and possible interaction or collision at some location in the heliosphere. The aforementioned uncertainties in the GCS fitted parameters are noted by inspecting the differences in the fitted values obtained from several independent attempts of applying GCS model to the CMEs.

\subsection{Estimation of kinematics in HI field-of-view}
\label{reconhi}
\begin{figure}[H]
\begin{center}
\includegraphics[scale=0.60]{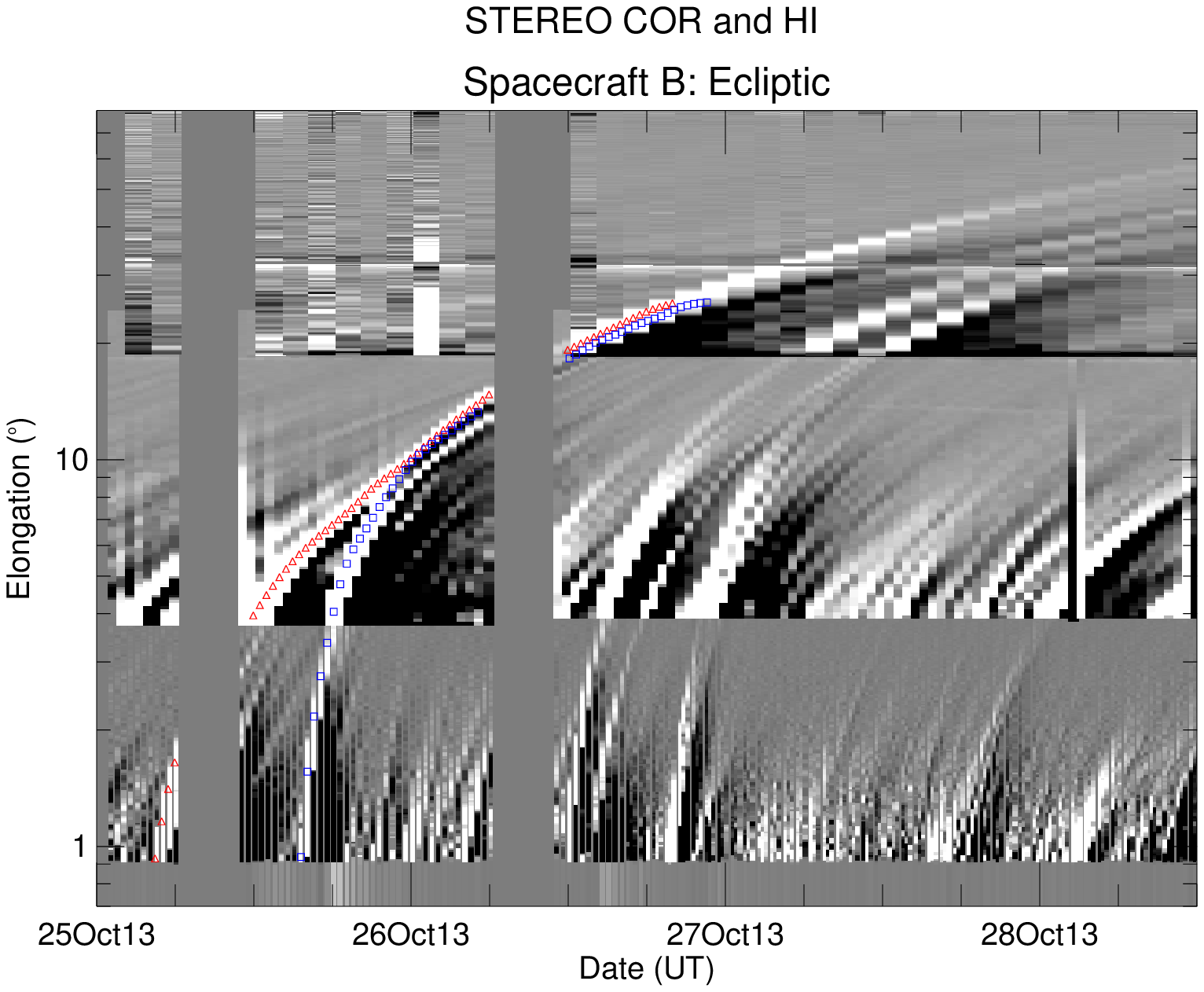}
\hspace{0.5cm}
\includegraphics[scale=0.60]{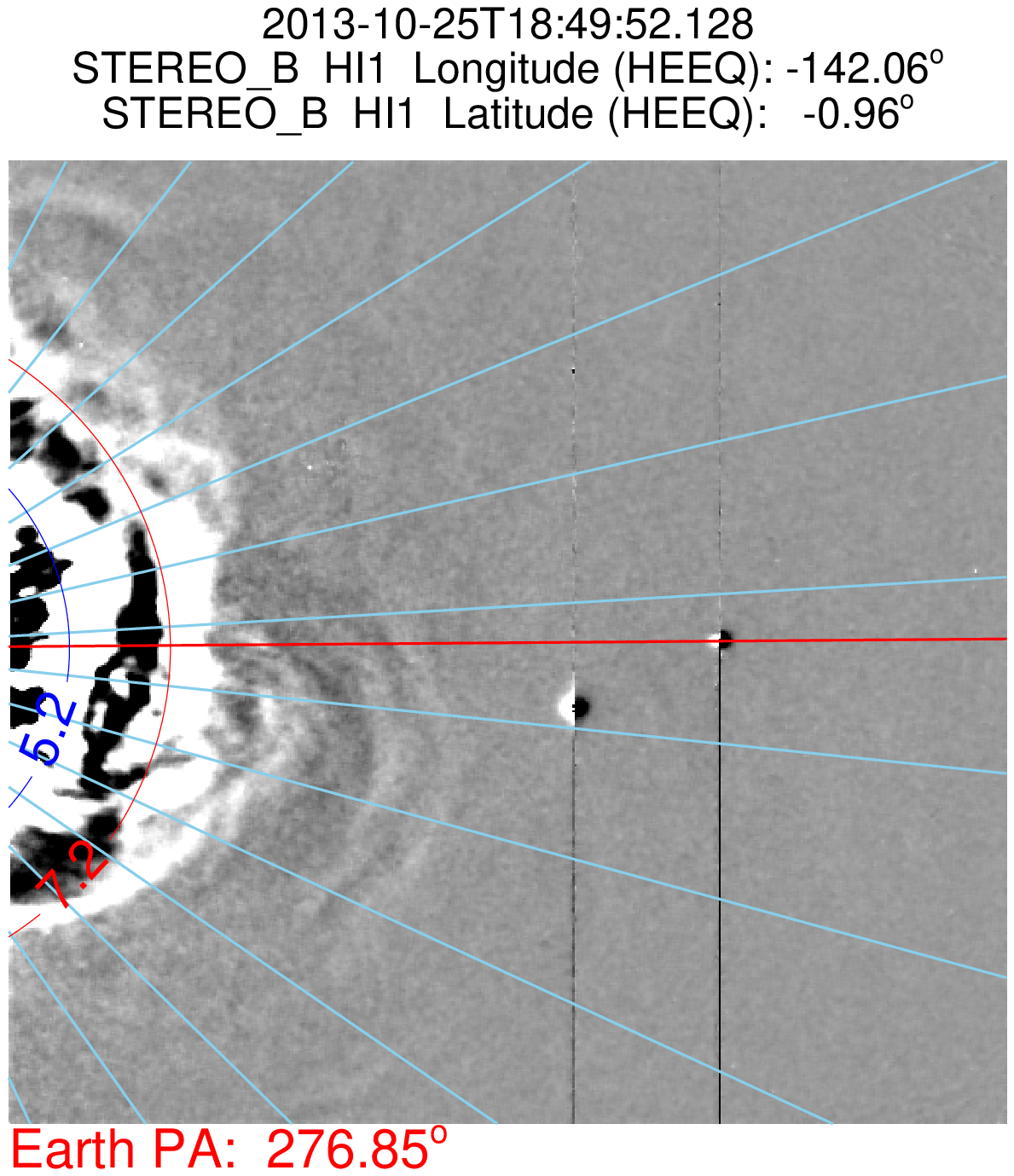}
\caption{Left panel shows the time-elongation plot (J-map) constructed using COR2-B and HI-B images of \textit{STEREO}/SECCHI Behind spacecraft for the period of October 25 to October 28 at 12:00 UT. The  symbol of $\triangle$ in red and $\square$ in blue mark the evolution of brightness enhancement (leading edge) due to CME1 and CME2, respectively. In right panel, the derived elongation of CMEs at 18:49 UT on October 25 is overplotted on HI1-B images.}
\label{jmap}
\end{center}
\end{figure}

\begin{figure}[H]
\begin{center}
\includegraphics[scale=0.95]{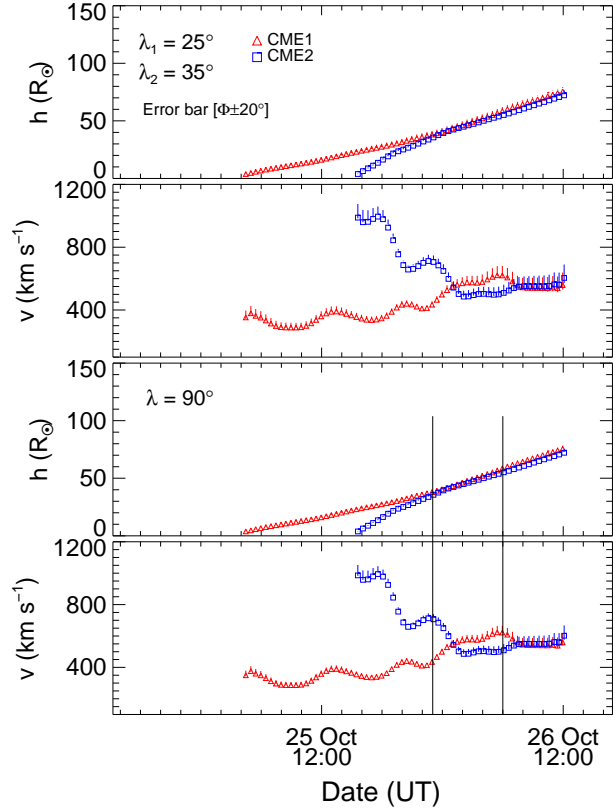}
\caption{From the top: the first and second panels show time variation of height and speed of the leading edge of the 2013 October 25 CMEs from SSE method. Third and fourth panels show the variation of height and speed from HM method. The vertical dashed lines in bottom panels mark the start and end of the collision phase. The small vertical line at each data point is the error bar due to an uncertainty of $\pm$20$^\circ$ considered in the value of propagation direction estimated from GCS model.}
\label{kinem}
\end{center}
\end{figure}

Examining the heliospheric evolution of the CMEs, we note that their leading front could not be observed in HI1-A field-of-view while their flank could remain visible only to a small elongation angle. This is because  of largely eastward propagation of the CMEs from the Sun-Earth line. Therefore, we used running difference images of COR2-B and HI-B to construct \textit{J}-map \citep{Sheeley1999,Davies2009} along the ecliptic  (left panel of Figure~\ref{jmap}). By manually clicking on the positively inclined bright features in the \textit{J}-map which correspond to enhancement of density due to the CMEs, we derived the elongation-time profile which are shown with red and blue in the figure. We overplotted the derived elongation profile of tracked features to the sequence of HI images and noted that the features correspond to the leading front of CMEs.  For exemplifying, the derived elongation angle of CMEs overplotted on a HI1-B image at an instant is shown in right panel of Figure~\ref{jmap}. We confirm that tracked leading edges of the CMEs meet around 10$^{\circ}$ elongation. Based on our earlier study on comparison of relative performance of reconstruction methods \citep{Mishra2014}, we implement SSE reconstruction method developed by \citet{Davies2012}. This method requires the propagation direction of CME and an input of cross-sectional angular half-width ($\lambda$) of CME fixed to an appropriate value. Based on the formulation described in Appendix of \citet{Mishra2015a}, we used the GCS fitted parameters and find the values of $\lambda$ for the CMEs. The calculated value of $\lambda$ for CME2 is around 35$\pm$5$^{\circ}$ which is almost 10$^{\circ}$ larger than its value for CME1. The value of $\lambda$ for CME1 and CME2 as of 25$^{\circ}$ and 35$^{\circ}$ is used as input  while implementing the SSE method. However, we acknowledge the errors in calculation of $\lambda$ and to examine its effect on the kinematics of the CMEs\citep{Mishra2015a}, we also implement Harmonic Mean (HM) method of \citet{Lugaz2009} which is equivalent to SSE method with $\lambda$=90$^{\circ}$. The derived elongation profile from \textit{J}-map is interpolated keeping half an hour interval to get kinematics data points closely connected. The estimated kinematics is shown in Figure~\ref{kinem}. The two upper and two lower panels in the figure show the kinematics from SSE and HM methods, respectively. The speed is estimated from adjacent distance points using a numerical differentiation with three-point Lagrangian interpolation and therefore have misleading nonphysical fluctuations. The smooth profile of speed can be derived by fitting the distance into a polynomial, but the information about short time variations in the speed during CME-CME interaction will also be lost. Despite having the nonphysical fluctuations in the speed, a careful inspection of the height-time tracks together with the speed profiles help to mark the timing for the CMEs getting in close contact of one another for the collision. 

Although we have taken extreme care in manual tracking, we acknowledge the possibility of error ($\approx$2$^{\circ}$) in elongation measurements from \textit{J}-map. Based on our earlier study in \citet{Mishra2015a}, we note that the error of around 2$^{\circ}$ in elongation has lesser effect on the kinematics than that of around 10$^{\circ}$ in propagation direction of the CMEs. The error in the estimated direction of the CMEs from the GCS method is around 5$^{\circ}$ in COR field-of-view. However the collision leading to a possibility of their real deflection \citep{Lugaz2012} may increase the error in direction in HI field-of-view where the collision takes place. The effect of real or artificial deflection \citep{Howard2009,Howard2011} of CMEs on the estimated kinematics may be crucial. To find the maximum possible error in kinematics induced from uncertainties in direction, a change of $\pm20^{\circ}$ in the propagation direction of the CMEs is considered in our study. We note that such a change in direction shows only a small effect on the kinematics at smaller elongation as has been reported earlier \citep{Wood2010,Howard2011,Mishra2014}. From the obtained kinematics, we notice that a change in $\lambda$ value for the selected CMEs has also a small effect on the kinematics obtained from SSE method. This supports the findings of earlier studies which have shown that change of $\lambda$ value significantly alters the kinematics of only those CMEs which are propagating more than 90$^{\circ}$ away from the Sun-observer line \citep{Liu2013,Mishra2015a,Vemareddy2015}. The kinematics at higher elongations are truncated where they have large fluctuations primarily because of large errors in tracking of the CMEs and hence elongation measurements from \textit{J-}map. Identifying the collision phase as described in \citet{Mishra2015a}, we notice that collision begins at October 25 23:00 UT and ends at October 26 06:00 UT. We emphasize that the timings for collision are based on the exchange of momentum between both the CMEs revealed from estimated kinematics of their leading edges. However, the leading edge of CME2 meets the trailing edge of CME1 before the observed time of acceleration of leading edge of CME1. As the signal transferring the momentum had to travel from the trailing edge of CME1 to its leading edge, the time for beginning of collision noted above is delayed than actual starting time of interaction between the CMEs. If the signal is carried by magnetohydrodynamical waves, the spatial length and plasma properties of CME1 will decide the time taken by the signal during its journey from the trailing edge to the leading edge of CME1. Thus the precise marking of start and end of momentum exchange between the CMEs is difficult. We note that during the collision phase of around 7 hr, CME1 accelerated from $u_{1}$=425 to $v_{1}$ =625 km s$^{-1}$ and CME2 decelerated from $u_{2}$=700 to $v_{2}=$500 km s$^{-1}$. The true masses of the both the CMEs are determined using COR2 images, following the method of \citet{Colaninno2009}. The mass of CME1 and CME2 are estimated to be 7.5 $\times$ 10$^{12}$ kg and 9.3 $\times$ 10$^{12}$ kg respectively. At the beginning of the collision, CME2 leading edge is around 37 $R_{\sun}$ while it is around 58 $R_{\sun}$ at the end of the collision phase.

\section{Coefficient of Restitution for the CMEs: Analysis and Results}\label{coli}
In this section, first we estimate the Newton's coefficient of restitution \citep{Newton1687} for the colliding CMEs following the treatment of head-on collision as described in \citep{Mishra2014a,Mishra2015}. Then we consider a more realistic scenario of oblique collision. As the selected CMEs are propagating along the same latitude, almost in the ecliptic along which \textit{J}-map is made, therefore Figure~\ref{kinem} represents the kinematics of only a portion of the CME moving in two-dimensional (2D) plane. This indicates that our consideration of head-on and oblique collision stand for 1D and 2D collision picture. Putting together the results of the analysis for 1D and 2D picture helps us to realize several interesting points which must be addressed for improving our understanding of collision nature of any magnetized and expanding plasma blobs, such as CMEs.

\begin{figure}
\begin{center}
\includegraphics[scale=0.72]{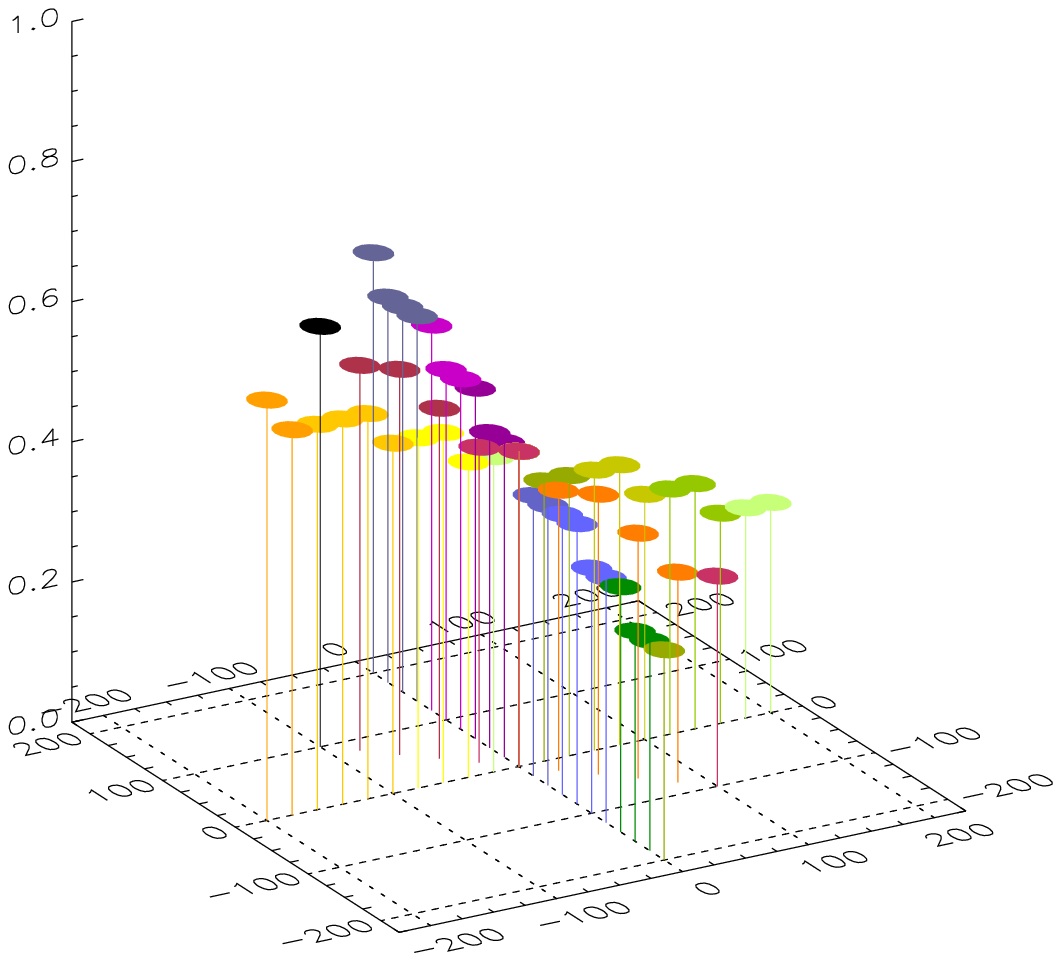}                                
\put (-70,0) {$X$}
\put (-195,15) {$Y$}
\put (-230,125) {$Z$}
\hspace{0.3cm}
\includegraphics[scale=0.72]{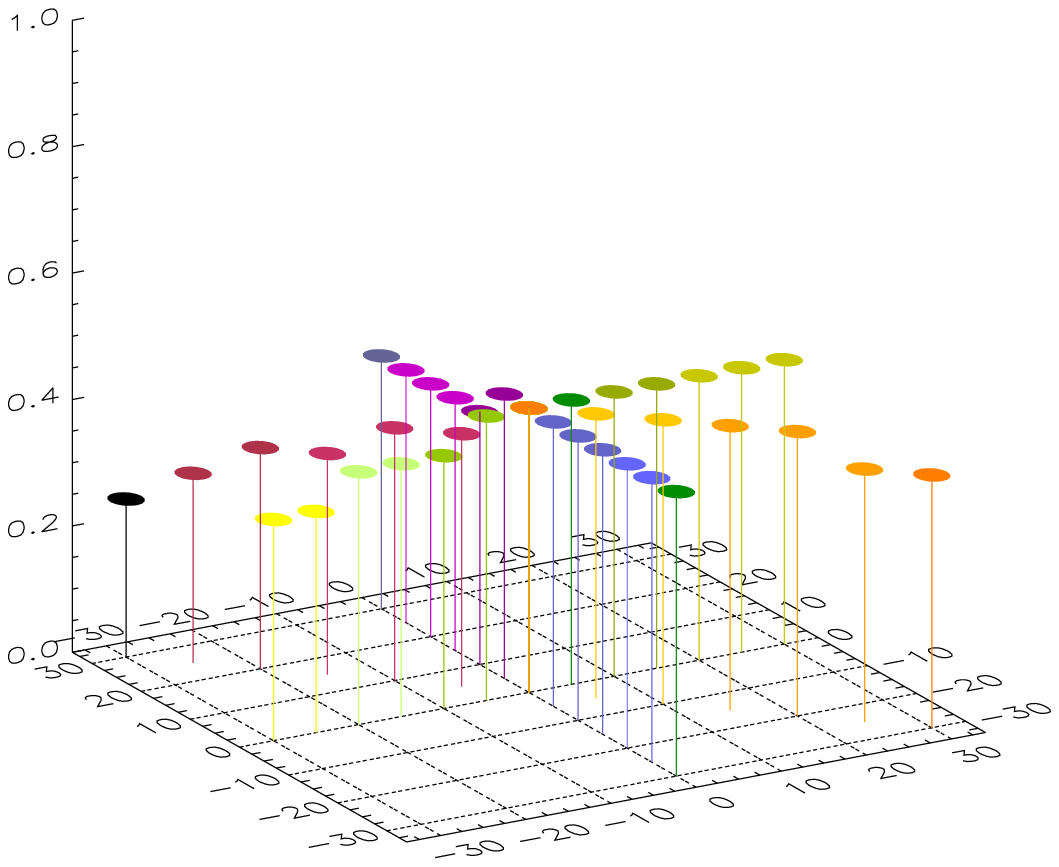} 
\put (-70,0) {$X$}
\put (-195,15) {$Y$}
\put (-230,125) {$Z$}\\
\vspace{0.1cm}
\includegraphics[scale=0.72]{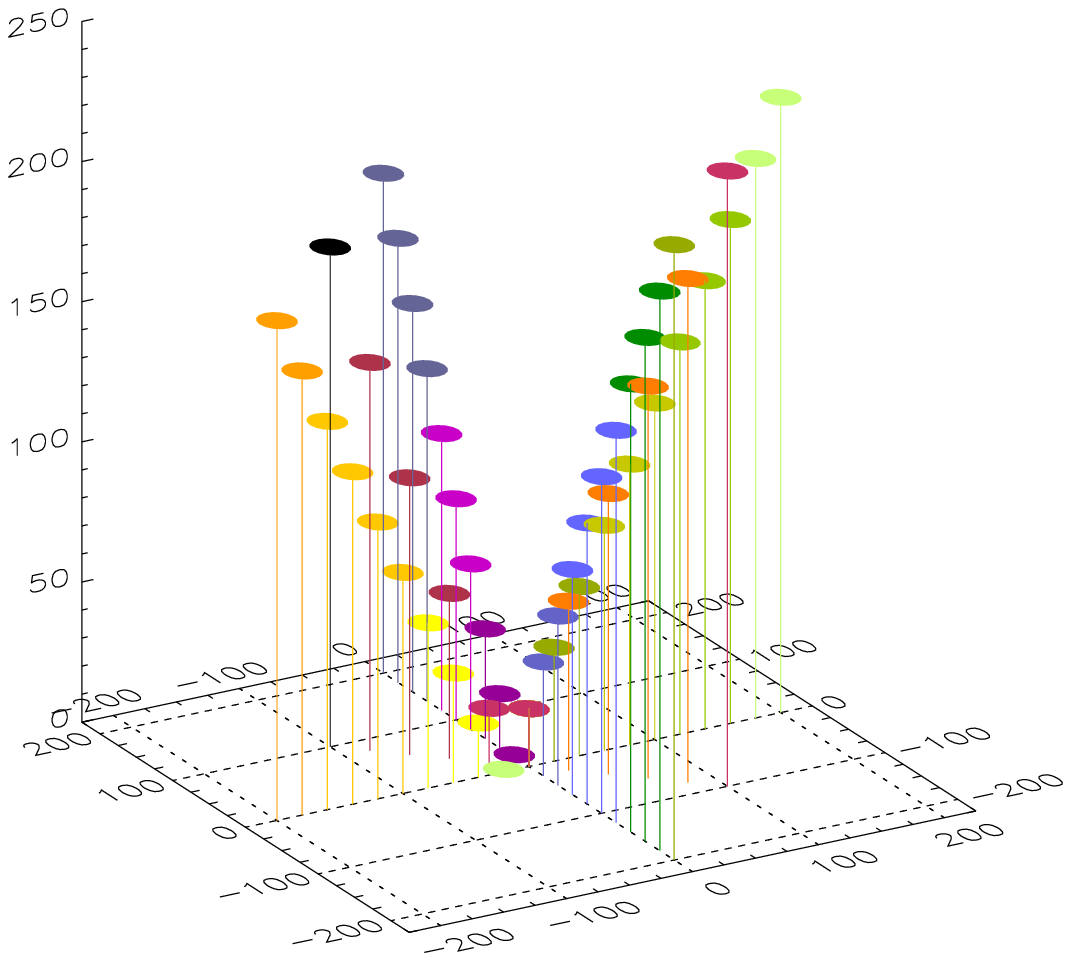} 
\put (-70,0) {$X$}
\put (-195,15) {$Y$}
\put (-230,125) {$Z$}
\hspace{0.3cm}
\includegraphics[scale=0.72]{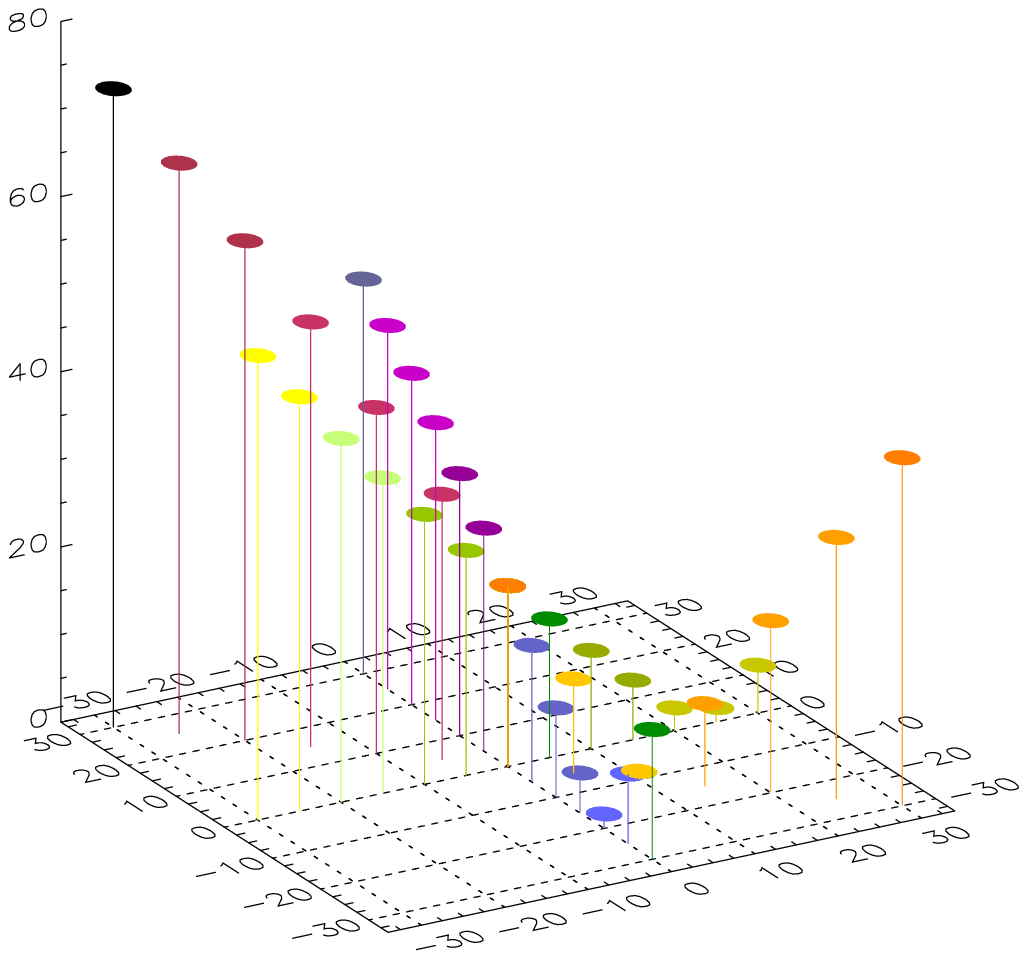} 
\put (-70,0) {$X$}
\put (-195,15) {$Y$}
\put (-230,125) {$Z$}\\
\vspace{0.1cm}
\includegraphics[scale=0.72]{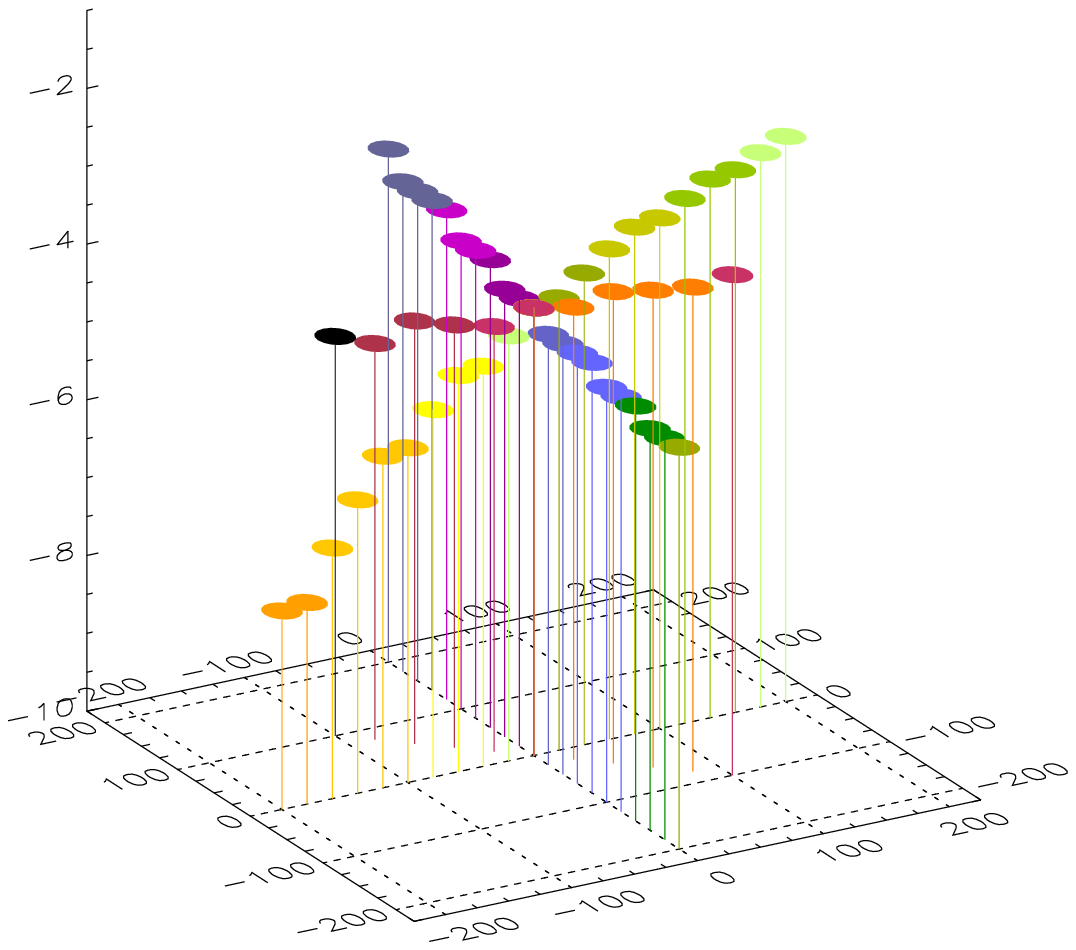} 
\put (-70,0) {$X$}
\put (-195,15) {$Y$}
\put (-230,125) {$Z$}
\hspace{0.3cm}
\includegraphics[scale=0.72]{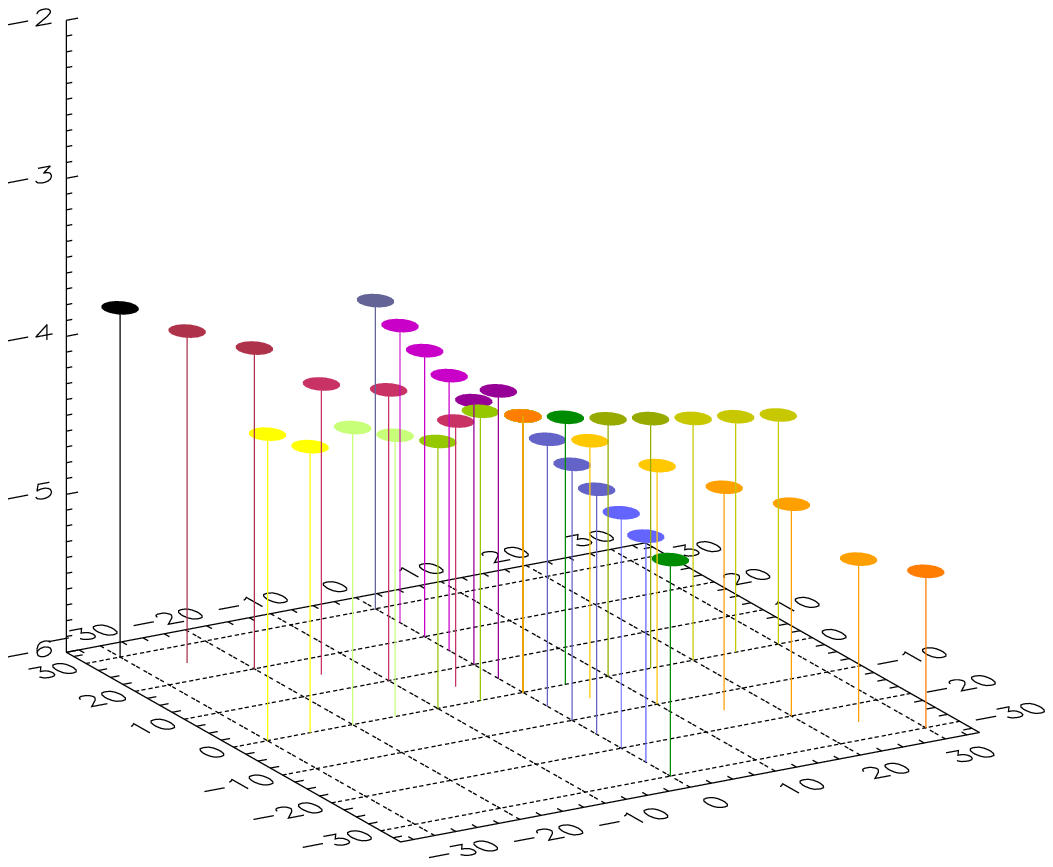} 
\put (-70,0) {$X$}
\put (-195,15) {$Y$}
\put (-230,125) {$Z$}
\caption{The $Z$-axis in top, middle and bottom panels stands for the parameters as coefficient of restitution ($e$), deviation in observed speed ($\sigma$) and change in total kinetic energy of the CMEs, respectively. In left panels: $X$-axis and $Y$-axis respectively show the uncertainties of $\pm$ 200 km s$^{-1}$ in the observed pre-collision speeds ($u_{1},u_{2}$) and post-collision ($v_{1},v_{2}$) of the CMEs. The different parameters shown on Z-axis in different panels with the same colors are corresponding to the equal uncertainties in observed speeds. In the right panels: the $X$ and $Y$-axis represents the $\pm$ 30\% uncertainties in estimated mass of the CME1 and CME2, respectively.  The different parameters shown on the Z-axis in different panels with the same color are corresponding to the equal uncertainties in the observed mass of the CMEs.} 
\label{res1D}
\end{center}
\end{figure}

\subsection{Head-on collision scenario} \label{1Dcol}

In this scenario, we assume that the speeds of both the CMEs before and after the collision are along the same line. We then modify the observed post-collision speeds ($v_{1}, v_{2}$) of the CMEs to satisfy the conservation law of momentum, and meanwhile minimize the deviation ($\sigma$) between the modified and observed post-collision speeds. We termed the modified speeds as theoretically estimated post-collision speeds ($v_{1th}, v_{2th}$) and define the deviation $\sigma = \sqrt{[(v_{1th} - v_{1})^{2} + (v_{2th} - v_{2})^{2}]/2}$. We find that the most optimized theoretically estimated post-collision speeds of CME1 and CME2 are 645 and 520 km s$^{-1}$, respectively, with $\sigma=20$ km s$^{-1}$. Thus, adopting the head-on collision and constraining the conservation of momentum, the coefficient of restitution 
($e$) is estimated to be about 0.45, suggesting an inelastic nature of collision. The total kinetic energy before the collision was 2.95 $\times$ 10$^{24}$ J, which decreased by $\approx$ 4.2\%  for the derived value of  $e$ =0.45.  The kinetic energy of CME1 and CME2 before the collision was 6.8 $\times$ 10$^{23}$ J and 2.27 $\times$ 10$^{24}$ J, respectively. After the collision, the kinetic energy of CME1 increased by 130\% while the kinetic energy of CME2 decreased by 44.5\% of its value before the collision.

There is a possibility of errors in estimated speed \citep{Lugaz2009,Davies2012} and the masses of the CMEs \citep{Colaninno2009}. Therefore, to examine their effect on the nature of collision we consider an arbitrary uncertainty in speed by $\pm$ 200 km s$^{-1}$ and in mass by 30\%. Without considering the expansion speeds of the CMEs, an uncertainty in their leading edge speeds is chosen such that they satisfy the 1D collision condition, i.e. $u_{2} \geq u_{1}, v_{1} \geq u_{1}$ and $v_{2} \leq u_{2}$. The results of estimated value of $e$, kinetic energy change and value of $\sigma$ is shown in Figure~\ref{res1D}. The $Z$-axis in top, middle and  bottom panels, represents the value of $e$, $\sigma$ and change in total kinetic energy of the CMEs. The $X$ and $Y$-axis in left panels represents the uncertainties in observed pre-collision and post-collision speeds of the CMEs while in right panels they represent the uncertainties of $\pm$30\% in measured mass of CME1 and CME2, respectively. From the figure, we can note that an error of $\pm$ 100 km s$^{-1}$ in observed speeds can result in the variation of coefficient of restitution from 0.3 to 0.6 with deviation ($\sigma$) values up to 200 km s$^{-1}$. We point out that while examining the effect of uncertainties, the speed of CMEs with uncertainties is now considered as observed speed and then their post-collision speed is modified for conservation of momentum with minimum value of deviation. Corresponding to an uncertainty of $\pm$ 100 km s$^{-1}$ in speeds, the total kinetic energy of the CMEs decreases by 3.5\% to 5.7\% of its value before the collision.   

The value of $e$ corresponding to larger uncertainties in speed and mass leading to higher value of $\sigma$ is less reliable. This is because large $\sigma$ suggests that the theoretically estimated observed speed satisfying the condition for conservation of momentum is significantly different than the observed speed of the CMEs. From the bottom panels, interestingly we note that  $\pm$30\% in the masses of the CMEs give a small change in value of $e$ from 0.25 to 0.45 with small value of $\sigma$ up to 75 km s$^{-1}$. Such findings suggest that even a large error in the mass estimates of the CMEs does not impose incertitude on our estimated $e$ value, i.e. nature of collision. The uncertainty in speeds may indirectly arise from the errors in direction, expansion speed and angular width of the CMEs. Hence the influence of these factors on the nature of the collision must be discussed.\\

\begin{figure}[!htb]
\begin{center}
\includegraphics[scale=0.6]{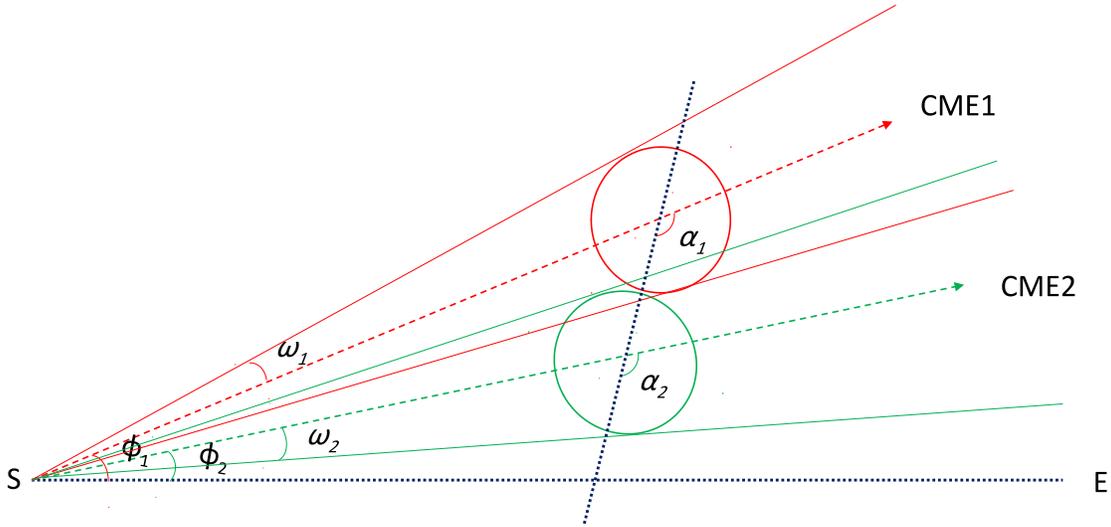}
\caption{The oblique collision of two CMEs assumed as spherical bubble is shown. The red and green circles represent the preceding CME1 and the following CME2. The dotted horizontal line marks the Sun-Earth line and the dotted oblique line is passing through the centroid of both the CMEs. The dashed red and green lines show the direction of propagation (i.e. longitude) of CME1 and CME2 as $\phi_{1}$ and $\phi_{2}$ from the Sun-Earth line, respectively. The $\omega_{1}$ and $\omega_{2}$ is the half angular width of the CME1 and CME2.  The $\alpha_{1}$ and $\alpha_{2}$ is the direction of propagation CME1 and CME2 from the line joining their centroid.}
\label{2Dpic}
\end{center}
\end{figure}
		
\subsection{Oblique collision scenario} \label{2Dcol}
For realistic situation close to the observations, we include possibility of oblique collision of the CMEs with their certain angular width. We consider CME1 and CME2 as expanding spherical bubbles propagating in $\phi_{1}$ and $\phi_{2}$ direction relative to a common reference line (i.e. Sun-Earth line in our case) with angular sizes of $\omega_{1}$ and $\omega_{2}$, respectively (Figure~\ref{2Dpic}). We consider that  the centroid of CME1 and CME2 is propagating in direction $\alpha_{1}$ and $\alpha_{2}$ relative to the line joining their centroids at the instant of collision.

Using reconstruction methods described in Section~\ref{reconcor}, we have estimated the $\phi_{1}$ and $\phi_{2}$ and now calculate the $\alpha_{1}$ and $\alpha_{2}$ by using the following relation. 

\begin{equation}
\label{eqalpha}
\begin{gathered}
\cos(|\phi_{1}-\phi_{2}|)\sin(\alpha_{1})+\sin(|\phi_{1}-\phi_{2}|)\cos(\alpha_{1})=\frac{\sin(|\phi_{1}-\phi_{2}|)-\sin(\omega_{2})\sin(\alpha_{1})}{\sin(\omega_{1})}\\
\alpha_{2}=\alpha_{1}+|\phi_{1}-\phi_{2}|
\end{gathered}
\end{equation}

We have estimated the speed of the leading edge of the CMEs, however speed of their centroid should be used to discuss the collision nature. Hence, the pre-collision speed of centroid for CME1 will be $u_{1c}=u_{1}-u_{1ex}$, where $u_{1}$ is the leading edge speed and $u_{1ex}$ is the expansion speed of CME1. By assuming that the CME expands in a way to keep its angular width as constant, we may get $u_{1ex}=u_{1}\sin(\omega_{1})/[1+\sin(\omega_{1})]$. Similarly the centroid speed of CME2 ($u_{2c}$) is equal to the difference between its leading edge speed ($u_{2}$) and expansion speed ($u_{2ex}$).  We consider that the post-collision direction of propagation of the CMEs relative to Sun-Earth line is $\phi_{1}^\prime$ and $\phi_{2}^\prime$, and relative to the line joining their centroid it is $\beta1$ and $\beta2$. The post-collision speed of centroid of CME1 and CME2 is $v_{1c}$ and $v_{2c}$, respectively. We note that due to the presence of the errors in observed pre- and post collision speeds they are not necessarily satisfying the momentum conservation. Also, during the collision if there is a deflection of the CMEs then their observed post-collision dynamics (Figure~\ref{kinem}) will be modified and its magnitude will also depend on the nature of collision. Hence, the observed post-collision dynamics cannot be used directly for studying the nature of collision. Therefore, we determined theoretically the post-collision speed ($v_{1cth}, v_{2cth}$) of the centroid of the CMEs using a certain value for $e$, which together allow the momentum to be conserved. Equations~\ref{eqvel} is used for the speed (i.e. $v_{cth}\cos(\beta)$) of the CMEs parallel to line joining their centroid. Under the collision scenario, an exchange of momentum takes place only along the line joining the centroids of the CMEs and therefore their speeds perpendicular to that line remains equal before and after the collision. 
Equation~\ref{eqprc} represents this condition mathematically for both the CMEs.

\begin{equation}
\label{eqvel}
\begin{gathered}
v_{1cth}\cos(\beta_{1})=\frac{m_{1}u_{1}\cos(\alpha_{1})+m_{2}u_{2}\cos(\alpha_{2})-m_{2}e[u_{1}\cos(\alpha_{1})-u_{2}\cos(\alpha_{2})]}{m_{1}+m_{2}}\\
v_{2cth}\cos(\beta_{2})=\frac{m_{1}u_{1}\cos(\alpha_{1})+m_{2}u_{2}\cos(\alpha_{2})+m_{1}e[u_{1}\cos(\alpha_{1})-u_{2}\cos(\alpha_{2})]}{m_{1}+m_{2}}
\end{gathered}
\end{equation}

 \begin{equation}
\label{eqprc}
\begin{gathered}
u_{1c}\sin(\alpha_{1})=v_{1cth}\sin(\beta_{1})\\
u_{2c}\sin(\alpha_{2})=v_{2cth}\sin(\beta_{2})
\end{gathered}
\end{equation}

Using equations~\ref{eqvel} and ~\ref{eqprc}, we determined the post-collision direction ($\beta_{1}$ \& $\beta_{2}$) of the CMEs and their post-collision speed ($v_{1cth}$ \& $v_{2cth}$) along this direction by putting a definite value of the coefficient of restitution ($e$). Considering that angular size of the CMEs remains unchanged before and after the collision, $v_{1cth},v_{2cth}$ is converted to post-collision leading edge speed ($v_{1th},v_{2th}$) which is compared with leading edge speed ($v_{1},v_{2}$) as observed using HI-1 data. We repeat the above mentioned procedures and calculate a set of theoretical values of final speed ($v_{1th}, v_{2th}$) corresponding to different values of $e$. The best suited value of $e$ is attributed to the nature of collision of the selected CMEs for which the deviation (i.e. $\sigma$ as defined in Section~\ref{1Dcol}) between the observed and the theoretically estimated post-collision leading edge speed is minimum. In our study, we have also estimated the post-collision direction of the CMEs $\phi_{1}^\prime$ and $\phi_{2}^\prime$ to measure the deflection of the colliding CMEs from the Sun-Earth line.

From the above description, it is clear that deflection of the CMEs during their collision has not been taken into consideration to derive their position and speed from SSE method. This is because the post-collision direction in our approach are found for a certain value of coefficient of restitution, and hence both are interrelated as we have five unknowns parameters and four equations (Equations~\ref{eqvel} and ~\ref{eqprc}) to deal with. By defining a parameter as variance ($\sigma$), we could manage to obtain the most likely value of $e$ and thus all five unknowns parameters. The value of the observed post-collision speeds from reconstruction methods (SSE and HM) are not directly used in our analysis. If the deflection of CMEs could have been estimated independently of any collision parameters, such as using elongation measurements and fitting methods \citep{Rouillard2008}, then we would have taken  this deflection into account while implementing SSE method for estimation of kinematics. However, we have used the estimated value of deflection for the CMEs to find their theoretical post-collision speeds. The theoretically estimated post-collision directions and speeds suggest that the deflection of the CMEs are indirectly taken into account for analyzing the collision picture. For the selected CMEs, we estimated the coefficient of restitution ($e$) in oblique collision scenario using the estimated kinematics and angular width of the CMEs (Section~\ref{reconhi}). From this, the value of  $e$ is estimated to be 0.6 for 2D collision scenario. This lead to 3.3\% decrease in the total kinetic energy of the CMEs. The value of $e$ from 2D scenario is almost equal to the value of $e$ estimated for in 1D collision scenario without taking uncertainties in the speed and mass of the CMEs. This finding is expected as the propagation direction of CME1 and CME2 is only 10$^{\circ}$ different to each other. Therefore, it will be interesting to see the value of $e$ in 2D scenario by taking the different propagation direction of the CMEs.    

\begin{figure}[H]
\begin{center}
\includegraphics[angle=90,scale=0.23]{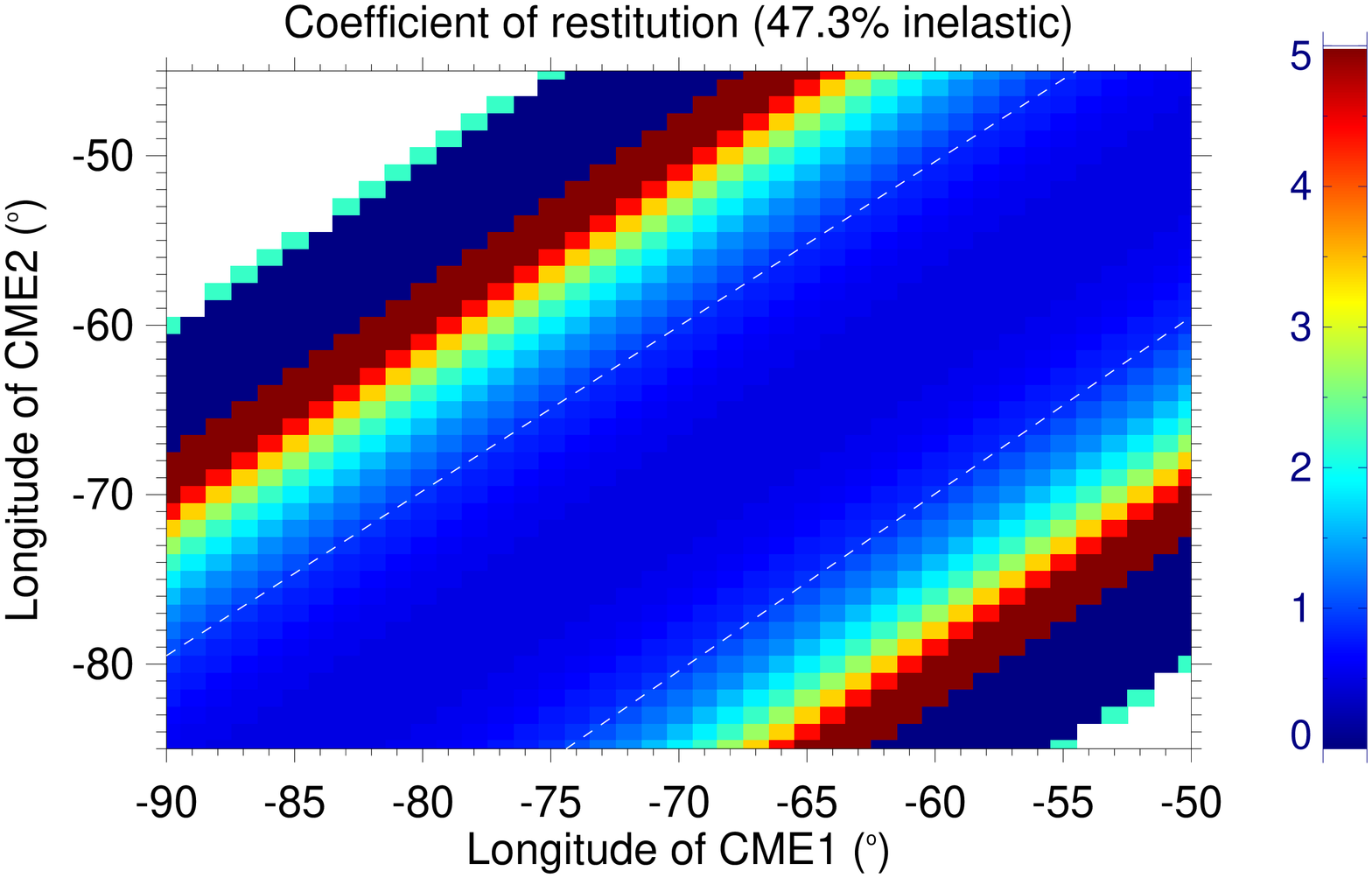}
\hspace{3mm}
\includegraphics[angle=90,scale=0.23]{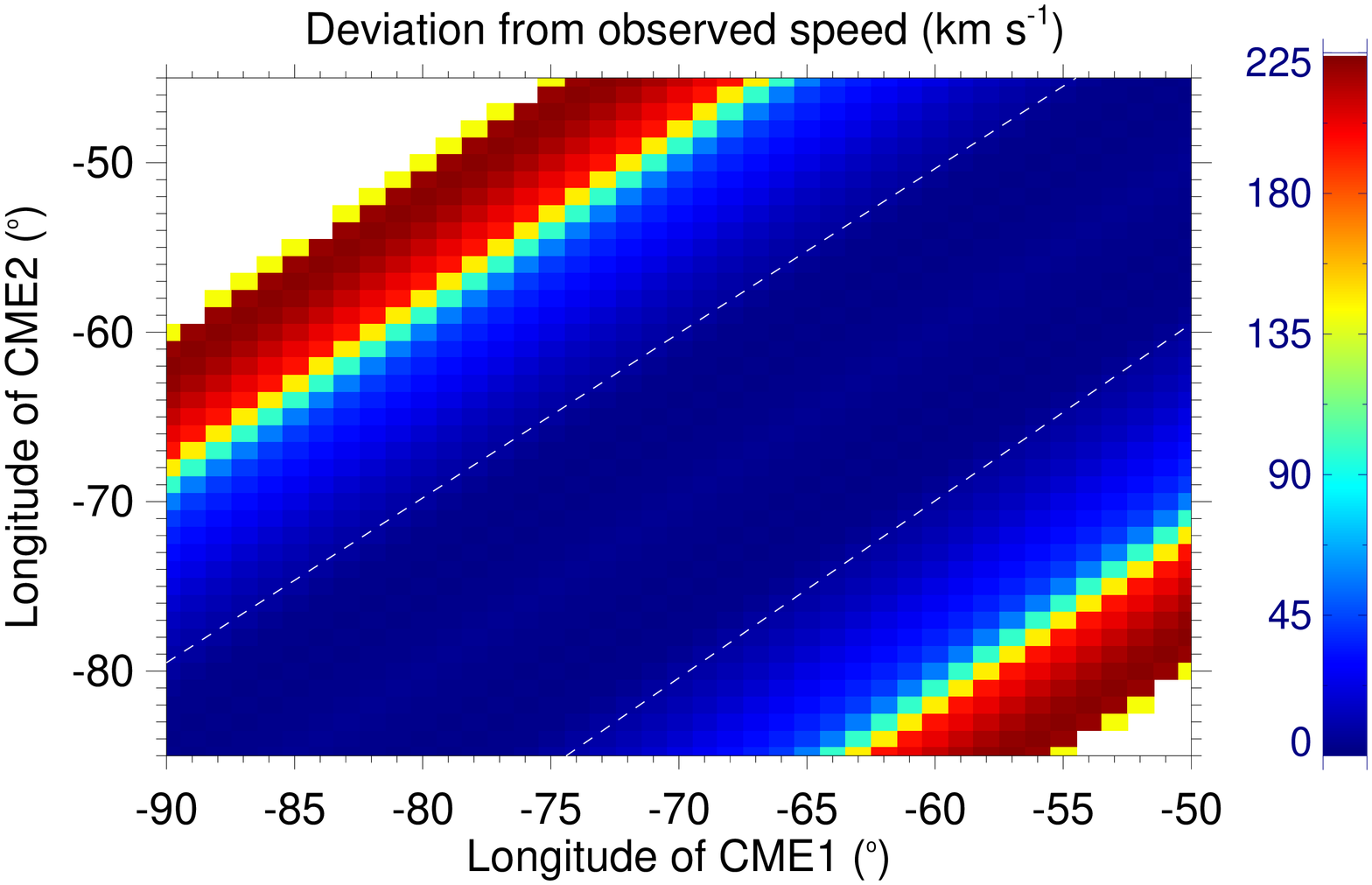}
\hspace{3mm}
\includegraphics[scale=0.30]{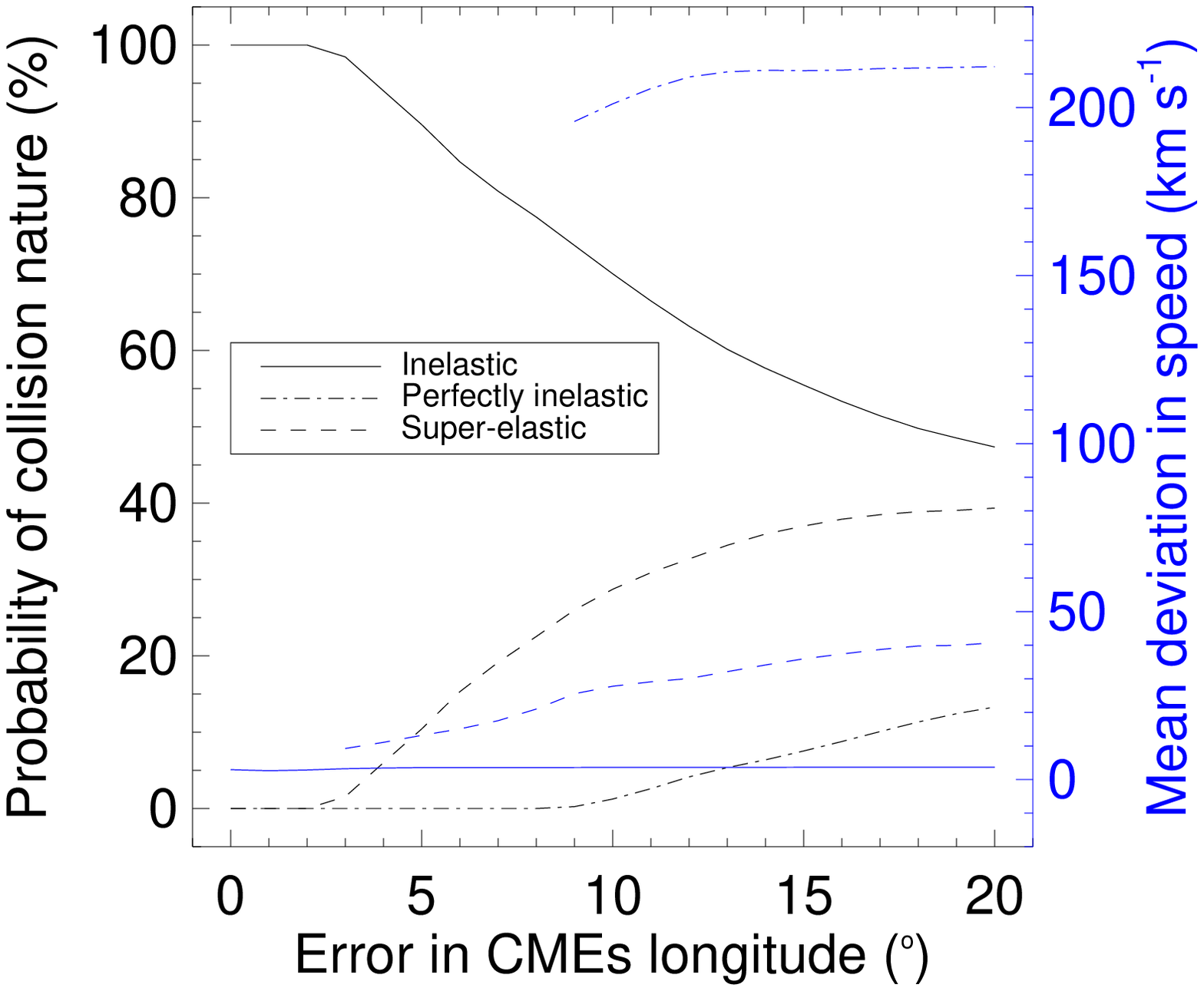}
\caption{Left and middle panels show the coefficient of restitution ($e$) and corresponding deviation ($\sigma$) between theoretical and observed post-collision speed values. The pre-collision longitudes of the CME1 and CME2 are shown on $X$ and $Y$-axis. The white dashed lines bound the region where nature of collision is inelastic, i.e. $e <$ 1. The color bar showing the range of the values shown in figures is also stacked. Right panel shows the probability of nature of collision and mean deviation in the speed with error in CMEs longitude.} 
\label{res2Ddir}
\end{center}
\end{figure}

\subsubsection{Effect of propagation direction}
\label{dirchng}
The collision takes place in HI field-of-view where a larger uncertainties in the direction is possible than that in COR field-of-view. To examine the role of direction in our analysis, we consider an uncertainty of $\pm$ 20$^{\circ}$ in the estimated longitude of CME1 and CME2 ($\phi_{1}$ and $\phi_{2}$) from GCS model described in Section~\ref{reconcor}. Then we followed the procedures described in Section~\ref{2Dcol} to calculate the coefficient of restitution and deviation in the speed which is shown in Figure~\ref{res2Ddir}. Only those pairs of direction for the CMEs are chosen for which the collision condition is satisfied. The condition is that the speed of leading edge of CME2 should be greater or equal to speed of trailing edge of CME1 along the line joining their centroids, i.e. [$u_{2c}\cos(\alpha_{2})+u_{2ex}] \geq [u_{1c}\cos(\alpha_{1})-u_{1ex}$], and separation angle between the CMEs should be lesser or equal to sum of their angular sizes, i.e. $ |\phi_{1}-\phi_{2}| \leq (\omega_{1}+\omega_{2 })$. The left and middle panel of the figure have missing values at the top left and bottom right corners where the longitudes of the CMEs do not satisfy the collision condition. Form the figure, we see that values of $e$ is less than unity in the region bounded between two white dashed lines. There are instances near the top left and bottom right corners where the value of $e$ is either equal to zero or greater than unity. The large $\sigma$ value at these corners corresponding to large separation angle between the CMEs suggests for lesser reliability of $e$ value there. Therefore, there is a higher probability of inelastic nature of collision for the 2013 October 25 CMEs.

In the right panel, we show the probability of the collision nature on the $Y$-axis in the left, and mean value of deviation in the speed on the $Y$-axis in the right, against the error in the CMEs longitudes on the $X$-axis. This shows that when error in longitude increases from $\pm$1 to $\pm$20$^{\circ}$, the probability of inelastic collision decreases from 100 to 47.3\% with mean deviation always less than 10 km s$^{-1}$. The probability is calculated based on the number of data points created using the pair of CMEs directions. The increasing errors in the longitude increase the probability of super-elastic nature of collision from 0 to 40\% and mean deviation in speed from 10 to 50 km s$^{-1}$. We also infer that the observed collision can never be attributed as perfectly inelastic as for this an extremely larger value of the mean deviation in speed is found. This emphasize the difference in results from head-on and oblique-collision scenario. We accept the limitations of our previous studies \citep{Mishra2014a,Mishra2015} where only head-on collision is adopted. We note that range of $e$ values for inelastic collision nature give the decrease in kinetic energy of the CMEs up to 4\% and super-elastic collision nature give an increase in the kinetic energy of the CMEs up to 15\% of its value before the collision.

It is noted that an error in the direction of the CMEs can result in different value of $e$ using our approach, probably because of the erroneously using the value of speed in our analysis. Hence, the estimated value of $e$ with larger value of $\sigma$ is reasonably lesser reliable. We consider the longitude of the CMEs propagating in east and west of the Sun-earth line with negative and positive sign respectively. The post-collision longitude of the CMEs will be $\phi^\prime=\phi + (\beta-\alpha$) for $\phi_{1}<\phi_{2}$ and $\phi^\prime=\phi-(\beta-\alpha)$ for $\phi_{1}>\phi_{2}$. We note that collision causes the deflection of both the CMEs up to $\pm$15$^{\circ}$ in the direction opposite to each other and therefore post-collision angular separation between the CMEs is larger than its pre-collision value. There is a probability of 80\% that deflection of CME2 is 0.5 to 0.7 times of CME1 deflection, and this is reasonable as CME2 is heavier than CME1. The deflection for the interacting CMEs of 2010 May 23-24 has been pointed out by \citet{Lugaz2012}.  From our analysis, it seems that the uncertainty in the directions of selected colliding CMEs have only pseudo-effect on the collision nature. We use the term `pseudo' because the use of different directions may cause the alteration in the value of $e$ which mistakenly would be believed if the larger value of $\sigma$ is overlooked. The larger value of $\sigma$ implies that the unreliable value of $e$ is due to using the kinematics which does not represent the observed collision picture. Such dubious effect on estimated $e$ value is possible because of the errors in $\phi$ and use of the observed speed estimated along different value of $\phi$. 

\begin{figure}[H]
\begin{center}
\includegraphics[angle=90,scale=0.37]{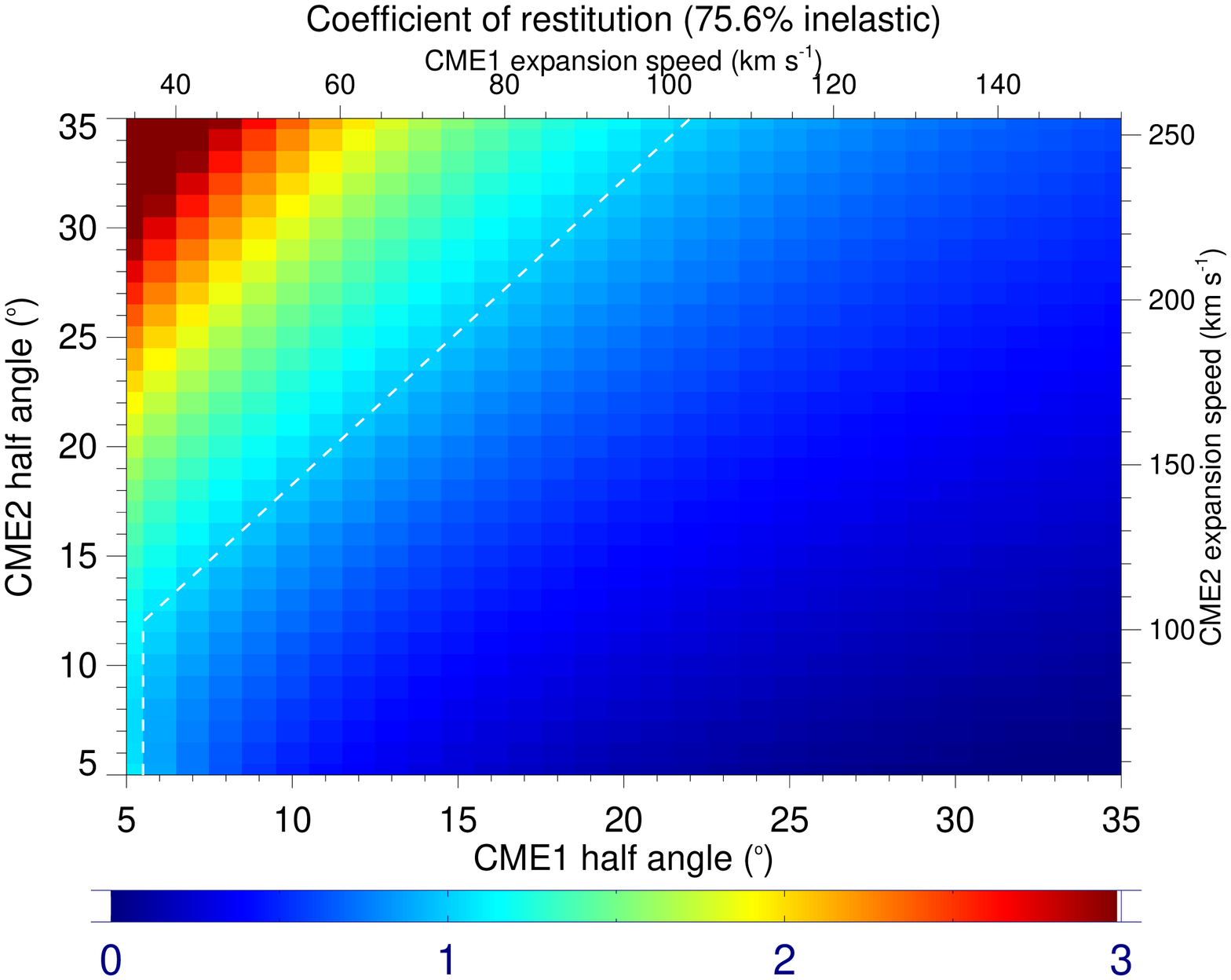}
\hspace{0.80cm}
\includegraphics[angle=90,scale=0.37]{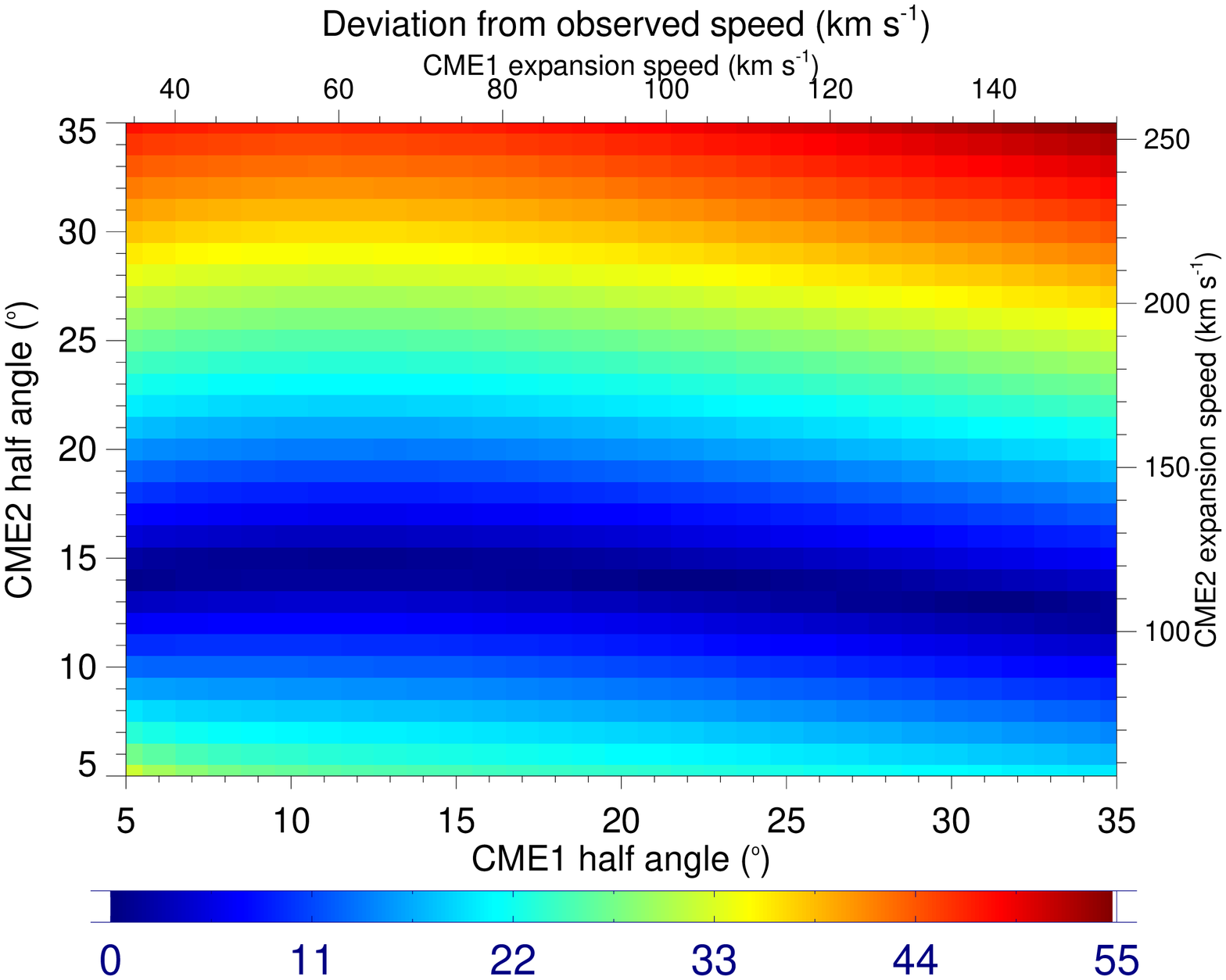}
\caption{Left and right panels show the coefficient of restitution ($e$) and corresponding deviation ($\sigma$) between theoretical and observed post-collision speed. Conventional $X$ and $Y$-axis represent the half angle ($\omega$) of the CME1 and CME2, respectively. The $X$ and $Y$-axis at top and right side, respectively show the expansion speed of CME1 and CME2. The dashed white line marks the boundary of super-elastic and inelastic regime. The color bar is stacked below the $X$-axis of the figure.} 
\label{res2Dwid}
\end{center}
\end{figure}

\subsubsection{Effect of Angular size}
\label{aschng}
We further examine the nature of collision of the CMEs because of uncertainty in their angular sizes. Large angular size of the CME directly implies the large value of expansion speed and therefore smaller speed of CME centroid on keeping its leading edge speed as constant. Keeping the kinematics same as estimated in Section~\ref{recon}, we arbitrary take the angular width ranging between 5 to 35$^{\circ}$ and repeat the procedures described in Section~\ref{2Dcol}. The estimated value of $e$ and $\sigma$ is shown in left and right panel of Figure~\ref{res2Dwid}. Despite having large uncertainty in the angular size, the probability of inelastic nature of collision is around 75.6\%. The deflection of both the CMEs is up to $\pm$10$^{\circ}$ in direction opposite to each other because of their collision. 

From the figure, it is clear that at the top left corner, the value of $e$ is greater than unity and represents for super-elastic nature of collision. The value of $e$ shown in Figure~\ref{res2Dwid} for super-elastic nature of collision corresponds to increase in kinetic energy of the CMEs up to 6\%, and for inelastic collision nature corresponds to decrease in kinetic energy up to 12\% of its value before the collision. Around 99\% data points for the super-elastic nature of collision show that the angular width of CME2 is more than 1.5 times of CME1 width.  Corresponding to this, the expansion speed of CME2 is more or equal to 2.0 times of CME1 expansion speed. The value of  $\sigma$ for the super-elastic collision nature ranges from 5 to 50 km s$^{-1}$ which is not larger than its value for inelastic nature of collision. The values of $e$ equal to zero at bottom-right corner of left panel of the figure are associated with  CME1 expansion speed which is more than 1.5 times of expansion speed of CME2. This suggests that super-elastic nature of collision is probable with larger expansion speed of the following CME.

\begin{figure}[H]
\begin{center}
\includegraphics[angle=90,scale=0.34]{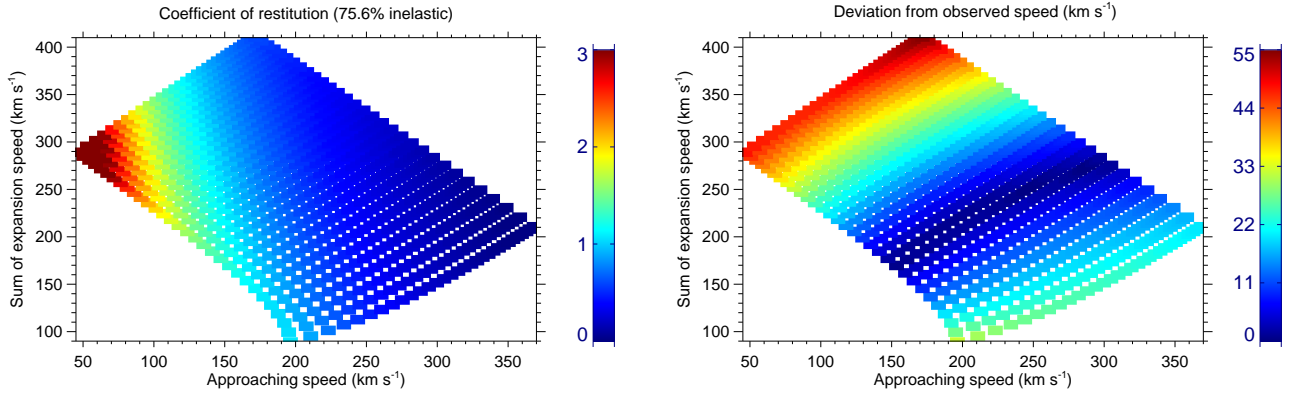}
\hspace{0.60cm}
\includegraphics[angle=90,scale=0.34]{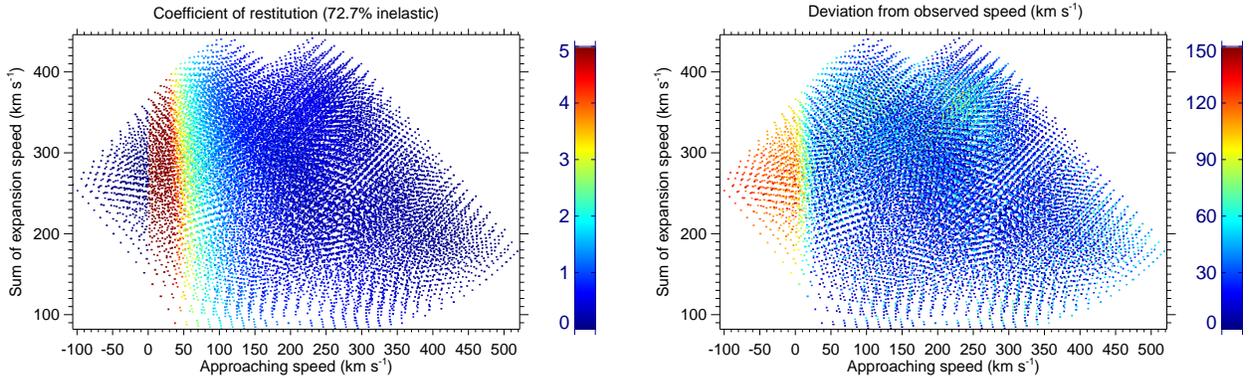}
\caption{Left and right panels show the coefficient of restitution ($e$) and corresponding deviation ($\sigma$) between theoretical and observed post-collision speed. $X$-axis represents the relative approaching speed of the CMEs, i.e. $[u_{2c}\cos(\alpha_{2})-u_{1c}\cos(\alpha_{1})]$.  The sum of expansion speed of CME1 and CME2, i.e. $u_{1ex}+u_{2ex}$, is shown along the $Y$-axis. The color bar is stacked right to the figure.} 
\label{res2Dwid_diff}
\end{center}
\end{figure}

Using the expansion speed corresponding to different angular sizes of the CMEs, we determined the speed of their centroid i.e. $u_{1c}$ and $u_{2c.}$. As per the suggestion made in \citet{Shen2012}, we examined the characteristic of collision with approaching speed of the CMEs ($|u_{2c}\cos(\alpha_{2})-u_{1c}\cos(\alpha_{1})|$) and sum of their expansion speed before the collision. Figure~\ref{res2Dwid_diff} (left) shows the variation in $e$ value against their relative approaching speed on $X$-axis and sum of expansion speeds at $Y$-axis. From the figure, it is clear that the nature of collision is found to be super-elastic when the sum of expansion speed is equal or larger than the relative approaching speed of two CMEs before the collision. This finding is consistent with the condition for super-elastic collision conceptualized in \citet{Shen2012,Shen2016}. However, this appears only as a necessary condition but not a sufficient condition for super-elastic collision. As among all the values of $e$ for inelastic collision nature around 63\% of them also satisfy this condition. For the selected CMEs under the assumed uncertainties in the half-angular width, we note that almost 99\% data points for inelastic nature of collision correspond to CME2 expansion speed ranging between around 0.3 to 2.5 times of CME1 expansion speed. In contrast, the expansion speed of CME2 is ranging between around 2 to 7 times of CME1 expansion speed for super elastic collision.

The value of $\sigma$ corresponding to the estimated value of $e$ is shown in Figure~\ref{res2Dwid_diff} (right). In this figure, unlike the case shown in Figure~\ref{res2Ddir}, we note that the value of $\sigma$ corresponding to super-elastic nature of collision is not larger than its value for inelastic collision nature. Therefore, the estimated value of $e$ for super-elastic collision is reliable in this case.  For the selected CMEs under the assumed uncertainties in their angular size, the ratio of CME2 to CME1 expansion speed as 2.0 or the low approaching speed around 150 km s$^{-1}$ is a threshold to turn on the super-elastic nature of collision. The larger expansion speed of CME2 than CME1, and lower approaching speed appear as two important condition for increasing the probability of super-elastic collision. Both conditions support each other as an increase in expansion speed of CME2 indirectly gives the lesser value of relative approaching speed of CME1 and CME2 than the sum of their expansion speed.

\subsubsection{Effect of initial speed and angular size}
\label{isaschng}
\begin{figure}[!htb]
\begin{center}
\includegraphics[angle=90,scale=0.33]{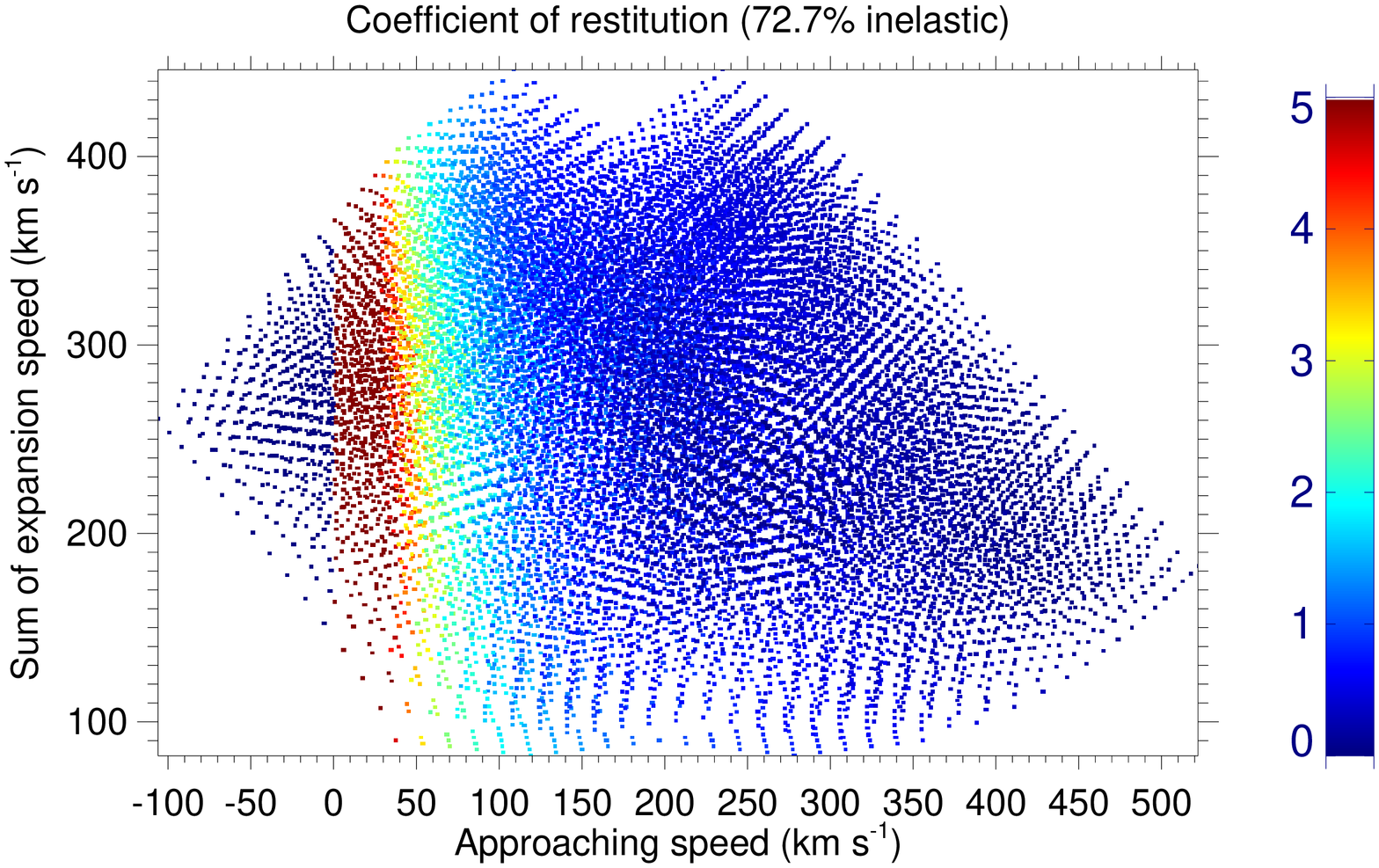}
\hspace{0.60cm}
\includegraphics[angle=90,scale=0.33]{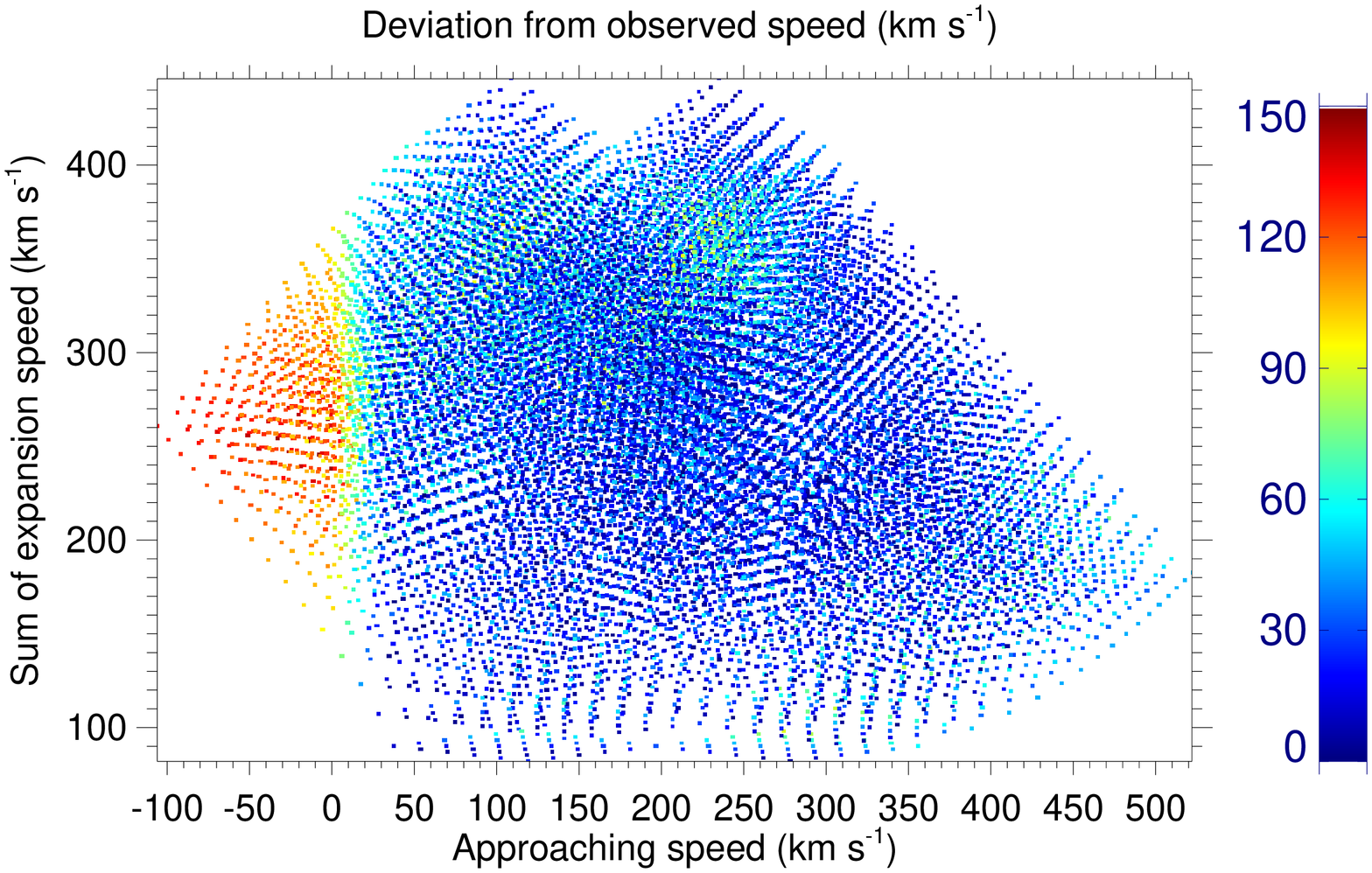}
\caption{Left and right panels show the coefficient of restitution ($e$) and corresponding deviation ($\sigma$) between theoretical and observed post-collision speed. $X$-axis represents the relative approaching speed of the CMEs, i.e. $[u_{2c}\cos(\alpha_{2})-u_{1c}\cos(\alpha_{1})]$.  The sum of expansion speed of CME1 and CME2, i.e. $u_{1ex}+u_{2ex}$, is shown along the $Y$-axis. The color bar is stacked right to the figure.} 
\label{res2Dwid_intspd_diff}
\end{center}
\end{figure}

We further consider the uncertainties of $\pm$ 100 km s$^{-1}$ in the observed pre-collision leading edge speed ($u_{1}$ and $u_{2}$) of the CMEs together with uncertainties in their angular width ($\omega_{1}$ and $\omega_{2}$). The longitude of the CME1 and CME2 is taken as estimated using GCS model of 3D reconstruction in Section~\ref{reconcor} and angular width of the CMEs is considered to range from 5 to 35$^{\circ}$. We estimated the value of coefficient of restitution ($e$) and deviation ($\sigma$) between theoretically derived and observed leading edge speed of the CMEs which is shown in Figure~\ref{res2Dwid_intspd_diff}. From this figure, we note the probability of 72.7\% for inelastic nature of collision. The probability of super-elastic collision is around 27\% and corresponds to an approaching speed ranging between 0 to 285 km s$^{-1}$. For these super-elastic collisions, the value of  $\sigma$ is not larger than its value for inelastic collision. Therefore, the values of $e>$1 are reliable. Among all the values of $e$ for super-elastic collision around 96\% of them correspond to larger values of the sum of expansion speed of two CMEs than their approaching speed before the collision. And around 98\% among these values of $e$ are associated with larger expansion speed of CME2 than CME1 expansion speed before the collision. We note that only around 45\% among all the values of $e<1$ have larger expansion speed of CME2 than CME1.  

We found that 1\% among all the values of $e<1$ correspond to ratio of CME2 to CME1 expansion speed around 5, however, around 12\% among all the values of $e>1$  satisfy this speed ratio. This suggest that as the ratio of expansion speed of CME2 to CME1 increases, the probability of super-elastic collision increases over inelastic one. The blank spaces at the corners in the images correspond to the values which do not satisfy the collision condition, as described in Section~\ref{2Dcol}. From the figure, the values of $e$ with negative approaching speed correspond to significantly larger value of $\sigma$ and therefore is not reliable. Negative approaching speed implies that collision of CMEs took place because of their larger expansion speed. Left panel of Figure~\ref{res2Dwid_intspd_diff} and ~\ref{res2Dwid_diff} shows that the decrease in the approaching speed increases the probability of super-elastic collision for the selected CMEs. This finding is consistent with the concept highlighted in \citet{Shen2016}. 

As the larger expansion speed and  internal pressure also imply to harder the CME, we ideate that collision tends to be super-elastic if a following CME is harder than preceding CME. Super-elastic nature of collision has been noticed in experiments where hard ceramic spheres impact a softer polycarbonate plate \citep{Louge2002,Kuninaka2004}. We suggest that internal pressure of a CME indirectly stands for physical nature of the macroscopic expanding plasma blobs. Therefore, the different physical characteristics of the CMEs plasma may give different nature of collision.

\section{Discussion}\label{Dis}
Our analysis shows that nature of collision of 2013 October 25 CMEs is inelastic. There also exist a smaller probability of super-elastic collision when uncertainties of angular width, expansion speed and initial speed of the CMEs is considered. Inelastic collision implies that some energy of the colliding CMEs is lost, instead of converting to kinetic energy. This may be due to deformation, compression, friction inside and over colliding surface of the system which contribute for conversion of some energy into heat. In our study, there are several idealistic assumptions which may no longer be valid in the real scenario. We have estimated the pre- and post-collision speed of the CMEs by marking the start and end of the collision phase. However, the exchange in the momentum of the CMEs might have started before the noticed collision \citep{Temmer2012,Temmer2015} because of change in the local environment for CME2, and driven shock by CME2 which may be passing through CME1. The role of shock in CME-CME interaction yet need to be understood properly \citep{Lugaz2005}. A fast-mode shock may dissipate its kinetic energy into thermal and magnetic energy. Thus ignoring the shock during the interaction probably makes the nature of the collision of two main bodies of the CMEs being under-estimated. Another limitation of our study is the lack of consideration of  momentum exchange between the CMEs and solar wind during the collision. For the collision of two relatively slow CMEs in \citet{Shen2012}, it has been shown that the acceleration of preceding CME due to solar wind was only about 6.5\% of that caused by the collision. In their study, the influence of solar wind on the following CME should be even smaller as its speed was closer to the solar wind speed than the preceding CME. For the case in our study, the speed of the preceding CME is close to the slow speed ambient solar wind, but the second CME was extremely fast. Thus, the difference in the pre- and post-collision speeds of the CMEs may not be completely attributed to the collision. Thus, the system of both the CMEs would not behave as a close system for which the conservation of momentum need to be be satisfied over the collision duration of few hr. This is also a reason why the theoretically estimated speed is not equal to measured post-collision speed for the CMEs.

Further, the identification of collision phase based on the estimated kinematics of only a portion of leading edges of the CMEs creates some uncertainty in our analysis. As per our definition of collision phase, the marked start time of collision is slightly postponed than the actual start time where the trailing edge of CME1 is hit by the leading edge of CME2. This causes the value of coefficient of restitution to be overestimated. The extent of overestimation depends on the deceleration of CME2 during the time of transporting the disturbance from trailing edge to leading edge of CME1. Based on the visual inspection of CMEs images, CME2 touches the trailing edge of CME1 around 18:00 UT on October 25. Thus, the marked start time of collision is postponed by 5 hr and CME2 leading edge speed is underestimated by 225 km s$^{-1}$. This finding roughly results in the overestimation of coefficient of restitution by 50\% in our study. On the other hand, the underestimation of coefficient of restitution is also inferred because of ignoring the contribution of shock in acceleration of CME1. It will be worth investigating the effect of these two source of errors causing competitively for the overestimation and underestimation of value of coefficient of restitution. Thus the difficulty in marking the collision phase introduces the errors into pre-and post-collision speed of the CMEs, however its effect on the estimated coefficient of restitution has not been considered rigorously in our study.

We have also estimated the mass of the CMEs in coronagraphic field-of-view and assume that it remains the same at collision site in HI1 field-of-view \citep{Carley2012,Bein2013}. However, \citet{Deforest2013} have considered the snowplow effect in the solar wind and shown that mass of a CME may increase by a factor of 2 to 3 in the heliosphere. Such an increase in the mass will change the magnitude of momentum exchange of the CMEs because of their interaction with solar wind which is ignored in our study. We further assume that the total mass of both the CMEs participates in the collision picture. In calculating the leading edge and expansion speed of the CME, we have assumed the CME as a circular structure in the ecliptic plane. This assumption may not necessarily be true but ease our analysis. Further, the angular width of this circular structure is determined based on the GCS modeled width and orientation of the flux rope of the CME. Our analysis of these selected CMEs quite matches the diagram (Figure 4d) in \citet{Shen2016}. The initial speed of the first CME is 485 km s$^{-1}$, which requires the initial approaching speed of the two CMEs to be less than 500 km s$^{-1}$ for a super-elastic process. However, the initial approaching speed of the two CMEs is about 515 km s$^{-1}$, higher than 500 km s$^{-1}$, so the collision is inelastic or has a greater probability to be inelastic.

In analysis presented in section~\ref{dirchng}, an arbitrary uncertainty in the direction is not directly used to estimate the change in the CMEs speed from SSE method. However, the speeds profiles from SSE method having the same elongation profiles is more or less independent of the direction at low elongation angle where collision occurs in our case \citep{Howard2011,Mishra2014}. This is also evident from the extremely small error bars in Figure~\ref{kinem}. The error in speed because of not using SSE method with a change in direction is much lesser than the possible errors in speed from other sources, such as deciding the boundary of collision phase, shock-CME interaction, and drag forces on the CMEs. The effect of errors in the speed is separately dealt in our study. Moreover, the uncertainties in direction of the CMEs is taken in the step of $\pm$1$^{\circ}$, hence repeating the SSE method around hundreds times appears impractical and futile in the context of the present study. However, we did not completely overlook the interrelatedness of direction and speed. As the change in direction is considered with change in the value of $\alpha1$ and $\alpha2$ in Equations~\ref{eqvel} which will eventually modify the originally observed speed to be taken for our analysis. Thus, indirectly we are taking into account the altered speed because of the change in the direction. Similarly, the deflected direction of the CMEs is not used in SSE method to derive their post-collision speeds. For the CMEs, a change in the direction would give a new elongation profiles for the observer. Thus the measured elongation angles and direction of propagation of CMEs are also linked. Since, we are using a fixed track of elongation from the \textit{J-}map, such effect of direction on the kinematics are difficult to explore in a true sense. Succinctly, our analysis represents the uncertainty in the results due to observational error for a single parameter at a time. Such analysis could be considered as case study for CMEs similar to the 2013 October CMEs having only one parameter different.

Our study have considered no change in the morphology and angular width of the CMEs during collision, and  only the linear speed of their centroid is used to calculate the change in their kinetic energy. As it is difficult to estimate the extent of compression and possible rotation or deflection of the CMEs during the collision. One of the major limitations of our study is that it completely ignores the magnetic field configuration of the CMEs. However, using numerical simulation \citet{Lugaz2013} have explored the role of relative orientation of magnetic flux rope in CMEs. We have considered uncertainty in individual parameter at a time, however uncertainty in several parameters need to be considered together to assess the reliability of estimated value of coefficient of restitution. We are unable to provide any information on the physical processes during the collision, however our study suggests that expansion speed of the CMEs play a role in deciding the nature of collision. The change in the angular width of the CMEs directly reflects the change in the expansion speed of the CME \citep{Gopalswamy2009}. The expansion speed of the CME is due to its larger internal pressure than ambient medium \citep{Wang2009}. Hence, we believe that a following CME having larger internal pressure leads to super-elastic collision if it hits the preceding CME with lower internal pressure. This may be due to the dissipation of magnetic and thermal pressure of the CME2 into kinetic energy. The change in the contact area of CMEs with different separation angles, i.e. different longitudes, may also partially contribute in deciding the nature of collision. However, with the error in direction leading to error in the speed, this could not be confirmed from our analysis. Several case studies of colliding CMEs, including the cases analyzed earlier \citep{Lugaz2012,Mishra2014a,Colaninno2015,Mishra2015}, are needed to investigate the role of duration of collision, contact area and their expansion speeds in converting the internal magnetic or thermal energy into the kinetic energy. We also note that considered uncertainties in angular width, speed and propagation direction of the CMEs also modify the distance of collision site. It is imperative to examine the affect of collision distance on the nature of colliding CMEs. For this, in another study, we are looking several cases of interacting CMEs wherein some are colliding close to and some away from the Sun.

There is a great scope of such studies towards the practical purpose of space weather forecasting. We think that the role of compression and subsequent expansion of CMEs, and thermodynamic changes inside the CMEs during collision phase must be well explored. We opine that elastic or super-elastic collision is not possible unless some physical processes during the collision convert magnetic or thermal energy of the CMEs to the kinetic energy. These processes such as magnetic reconnection may be crucial in increasing the post-collision macroscopic dynamics of the CMEs. In our earlier study \citep{Mishra2014a}, we have shown possible signatures of magnetic reconnection in the in-situ observations as a result of CME-CME interaction. Therefore, the role of magnetic reconnection in producing additional kinetic energy to the system and in deciding the type of in-situ structure of both CMEs need to be investigated. Thus, the relative orientation of magnetic flux rope of the interacting CMEs may influence the the nature of collision. The current study also highlights the drawbacks of our earlier study \citep{Mishra2014a,Mishra2015} where expansion speed of the CME was not taken into account and a simplistic scenario of head-on collision was adopted. Although, the coefficient of restitution estimated for the CMEs as per the Newton's definition seems to be a fairly reasonable approach. There are three definitions for coefficient of restitution by Newton (kinematic), Poisson (kinetic)  and Stronge (energetic) \citep{Brach1984,Stewart2000,Lubarda2010}, we need to contemplate as to which definition is more suitable for the observed CMEs in real scenario. We also plan to perform magnetohydrodynamics simulation for this interacting CMEs following the approach described in \citet{Shen2013,Shen2016} to examine the consistency between observation and simulation based study. \\

\section{Conclusion}\label{Res}
We have made an attempt to understand the uncertainties in the nature of collision of magnetized expanding plasma blobs by analyzing the interacting CMEs of 2013 October 25. Our analysis suggests for inelastic nature of collision for the selected CMEs. Uncertainties in the collision nature due to the error in direction, mass, angular width, expansion and propagation speed is examined. We show that the mass of the CMEs has almost no effect on deciding their nature of collision. Similar results have also been presented by \citet{Shen2012}. We note that that the head-on collision scenario causes the $e$ value to be underestimated than that of oblique collision. For the selected CMEs, the probability of inelastic nature of collision decreases with increasing the errors in the longitude of the CMEs. The values of $e>1$ corresponding to larger errors in longitudes leads to larger inconsistency with observed dynamics of the CMEs and therefore seems unreliable. This made us to acknowledge the pseudo effect of propagation direction of the CMEs on their collision nature. To estimate a reliable value of $e$, we emphasize that  error in the CMEs direction should be considered along with the errors in CME dynamics. Our analysis in a scenario of oblique collision clearly finds that deflection of interacting CMEs is an inevitable phenomenon.

The observed kinematics of the CMEs and their angular half-width ranging between 5 to 35$^{\circ}$ results in probability of around 75.6\% for inelastic nature of collision. The collision nature is found to be super-elastic when ratio of CME2 to CME1 angular half-width is greater or equal to 1.5.  We also noted super-elastic collision when the expansion speed of CME2 is greater or equal to 2 times of expansion speed of CME1. Our study finds that the lower approaching speed of the CMEs results in a greater probability of super-elastic collision. Further, an uncertainty of 100 km s$^{-1}$ in the initial speed of the CMEs together with the variation of their angular half-width from 5 to 35$^{\circ}$ lead to probability of 72.7\% for inelastic nature of collision. From our analysis, we establish a concept that the larger expansion speed of CME2 than CME1, and larger values of their sum over CMEs approaching speed tend to increase the probability of super-elastic collision \citep{Shen2012,Shen2016}. We conclude that if the expansion speed of following CME2 is larger than preceding CME1, it gives relatively low approaching speed before the collision and relatively high separation speed after the collision causing the nature of collision to be super-elastic. From our analysis for the CMEs of 2013 October, the relative expansion speed of the CMEs appears as a strong factor than relative approaching speed for deciding the nature of collision. Further study is needed to clearly understand the sufficient condition for inelastic or super-elastic collision. \\

We acknowledge UK Solar System Data Center for providing the processed Level-2 \textit{STEREO}/HI data.  The work is supported by the NSFC grant Nos. 41131065, 41574165 and 41421063. We also thank the reviewer whose comments have greatly improved this paper. W.M. is supported by the Chinese Academy of Sciences (CAS) President’s International Fellowship Initiative (PIFI) grant No. 2015PE015. 
\newpage


\begin{thebibliography}{}
\expandafter\ifx\csname natexlab\endcsname\relax\def\natexlab#1{#1}\fi

\bibitem[{{Bein} {et~al.}(2013){Bein}, {Temmer}, {Vourlidas}, {Veronig}, \&
  {Utz}}]{Bein2013}
{Bein}, B.~M., {Temmer}, M., {Vourlidas}, A., {Veronig}, A.~M., \& {Utz}, D.
  2013, \apj, 768, 31

\bibitem[{{Brach}(1984)}]{Brach1984}
{Brach}, R.~M. 1984, Journal of Applied Mechanics, 51, 164

\bibitem[{{Brueckner} {et~al.}(1995){Brueckner}, {Howard}, {Koomen},
  {Korendyke}, {Michels}, {Moses}, {Socker}, {Dere}, {Lamy}, {Llebaria},
  {Bout}, {Schwenn}, {Simnett}, {Bedford}, \& {Eyles}}]{Brueckner1995}
{Brueckner}, G.~E., {Howard}, R.~A., {Koomen}, M.~J., {et~al.} 1995, \solphys,
  162, 357

\bibitem[{{Burlaga} {et~al.}(1987){Burlaga}, {Behannon}, \&
  {Klein}}]{Burlaga1987}
{Burlaga}, L.~F., {Behannon}, K.~W., \& {Klein}, L.~W. 1987, \jgr, 92, 5725

\bibitem[{{Carley} {et~al.}(2012){Carley}, {McAteer}, \&
  {Gallagher}}]{Carley2012}
{Carley}, E.~P., {McAteer}, R.~T.~J., \& {Gallagher}, P.~T. 2012, \apj, 752, 36

\bibitem[{{Colaninno} \& {Vourlidas}(2009)}]{Colaninno2009}
{Colaninno}, R.~C., \& {Vourlidas}, A. 2009, \apj, 698, 852

\bibitem[{{Colaninno} \& {Vourlidas}(2015)}]{Colaninno2015}
{Colaninno}, R.~C., \& {Vourlidas}, A. 2015, \apj, 815, 70

\bibitem[{{Davies} {et~al.}(2013){Davies}, {Perry}, {Trines}, {Harrison},
  {Lugaz}, {M{\"o}stl}, {Liu}, \& {Steed}}]{Davies2013}
{Davies}, J.~A., {Perry}, C.~H., {Trines}, R.~M.~G.~M., {et~al.} 2013, \apj,
  776, 1

\bibitem[{{Davies} {et~al.}(2009){Davies}, {Harrison}, {Rouillard}, {Sheeley},
  {Perry}, {Bewsher}, {Davis}, {Eyles}, {Crothers}, \& {Brown}}]{Davies2009}
{Davies}, J.~A., {Harrison}, R.~A., {Rouillard}, A.~P., {et~al.} 2009, \grl,
  36, 2102

\bibitem[{{Davies} {et~al.}(2012){Davies}, {Harrison}, {Perry}, {M{\"o}stl},
  {Lugaz}, {Rollett}, {Davis}, {Crothers}, {Temmer}, {Eyles}, \&
  {Savani}}]{Davies2012}
{Davies}, J.~A., {Harrison}, R.~A., {Perry}, C.~H., {et~al.} 2012, \apj, 750,
  23

\bibitem[{{DeForest} {et~al.}(2013){DeForest}, {Howard}, \&
  {McComas}}]{Deforest2013}
{DeForest}, C.~E., {Howard}, T.~A., \& {McComas}, D.~J. 2013, \apj, 769, 43

\bibitem[{{Ding} {et~al.}(2014){Ding}, {Li}, {Jiang}, {Le}, {Shen}, {Wang},
  {Chen}, {Xu}, {Gu}, \& {Zhang}}]{Ding2014}
{Ding}, L.-G., {Li}, G., {Jiang}, Y., {et~al.} 2014, \apjl, 793, L35

\bibitem[{{Dungey}(1961)}]{Dungey1961}
{Dungey}, J.~W. 1961, \prl, 6, 47

\bibitem[{{Farrugia} \& {Berdichevsky}(2004)}]{Farrugia2004}
{Farrugia}, C., \& {Berdichevsky}, D. 2004, \ag, 22, 3679

\bibitem[{{Farrugia} {et~al.}(2006){Farrugia}, {Jordanova}, {Thomsen}, {Lu},
  {Cowley}, \& {Ogilvie}}]{Farrugia2006}
{Farrugia}, C.~J., {Jordanova}, V.~K., {Thomsen}, M.~F., {et~al.} 2006, \jgr,
  111, 11104

\bibitem[{{Gonzalez} {et~al.}(1994){Gonzalez}, {Joselyn}, {Kamide}, {Kroehl},
  {Rostoker}, {Tsurutani}, \& {Vasyliunas}}]{Gonzalez1994}
{Gonzalez}, W.~D., {Joselyn}, J.~A., {Kamide}, Y., {et~al.} 1994, \jgr, 99,
  5771

\bibitem[{{Gonzalez-Esparza} {et~al.}(2004){Gonzalez-Esparza}, {Santill{\'a}n},
  \& {Ferrer}}]{Gonzalez-Esparza2004}
{Gonzalez-Esparza}, A., {Santill{\'a}n}, A., \& {Ferrer}, J. 2004, Annales
  Geophysicae, 22, 3741

\bibitem[{{Gopalswamy} {et~al.}(2009){Gopalswamy}, {Dal Lago}, {Yashiro}, \&
  {Akiyama}}]{Gopalswamy2009}
{Gopalswamy}, N., {Dal Lago}, A., {Yashiro}, S., \& {Akiyama}, S. 2009, Central
  European Astrophysical Bulletin, 33, 115

\bibitem[{{Gopalswamy} {et~al.}(2001){Gopalswamy}, {Yashiro}, {Kaiser},
  {Howard}, \& {Bougeret}}]{Gopalswamy2001apj}
{Gopalswamy}, N., {Yashiro}, S., {Kaiser}, M.~L., {Howard}, R.~A., \&
  {Bougeret}, J.-L. 2001, \apjl, 548, L91

\bibitem[{{Gosling}(1993)}]{Gosling1993}
{Gosling}, J.~T. 1993, \jgr, 98, 18937

\bibitem[{{Harrison} {et~al.}(2012){Harrison}, {Davies}, {M{\"o}stl}, {Liu},
  {Temmer}, {Bisi}, {Eastwood}, {de Koning}, {Nitta}, {Rollett}, {Farrugia},
  {Forsyth}, {Jackson}, {Jensen}, {Kilpua}, {Odstrcil}, \&
  {Webb}}]{Harrison2012}
{Harrison}, R.~A., {Davies}, J.~A., {M{\"o}stl}, C., {et~al.} 2012, \apj, 750,
  45

\bibitem[{{Howard}(2011)}]{Howard2011}
{Howard}, T.~A. 2011, \jastp, 73, 1242

\bibitem[{{Howard} \& {Tappin}(2009)}]{Howard2009}
{Howard}, T.~A., \& {Tappin}, S.~J. 2009, \ssr, 147, 31

\bibitem[{{Intriligator}(1976)}]{Intriligator1976}
{Intriligator}, D.~S. 1976, \ssr, 19, 629

\bibitem[{{Kaiser} {et~al.}(2008){Kaiser}, {Kucera}, {Davila}, {St.~Cyr},
  {Guhathakurta}, \& {Christian}}]{Kaiser2008}
{Kaiser}, M.~L., {Kucera}, T.~A., {Davila}, J.~M., {et~al.} 2008, \ssr, 136, 5

\bibitem[{{Kuninaka} \& {Hayakawa}(2004)}]{Kuninaka2004}
{Kuninaka}, H., \& {Hayakawa}, H. 2004, Physical Review Letters, 93, 154301

\bibitem[{{Liu} {et~al.}(2013){Liu}, {Luhmann}, {Lugaz}, {M{\"o}stl}, {Davies},
  {Bale}, \& {Lin}}]{Liu2013}
{Liu}, Y.~D., {Luhmann}, J.~G., {Lugaz}, N., {et~al.} 2013, \apj, 769, 45

\bibitem[{{Liu} {et~al.}(2012){Liu}, {Luhmann}, {M{\"o}stl},
  {Martinez-Oliveros}, {Bale}, {Lin}, {Harrison}, {Temmer}, {Webb}, \&
  {Odstrcil}}]{Liu2012}
{Liu}, Y.~D., {Luhmann}, J.~G., {M{\"o}stl}, C., {et~al.} 2012, \apjl, 746, L15

\bibitem[{{Louge} \& {Adams}(2002)}]{Louge2002}
{Louge}, M.~Y., \& {Adams}, M.~E. 2002, \pre, 65, 021303

\bibitem[{{Lubarda}(2010)}]{Lubarda2010}
{Lubarda}, V.~A. 2010, Journal of Applied Mechanics, 77, 011006

\bibitem[{{Lugaz} \& {Farrugia}(2014)}]{Lugaz2014}
{Lugaz}, N., \& {Farrugia}, C.~J. 2014, \grl, 41, 769

\bibitem[{{Lugaz} {et~al.}(2012){Lugaz}, {Farrugia}, {Davies}, {M{\"o}stl},
  {Davis}, {Roussev}, \& {Temmer}}]{Lugaz2012}
{Lugaz}, N., {Farrugia}, C.~J., {Davies}, J.~A., {et~al.} 2012, \apj, 759, 68

\bibitem[{{Lugaz} {et~al.}(2013){Lugaz}, {Farrugia}, {Manchester}, \&
  {Schwadron}}]{Lugaz2013}
{Lugaz}, N., {Farrugia}, C.~J., {Manchester}, IV, W.~B., \& {Schwadron}, N.
  2013, \apj, 778, 20

\bibitem[{{Lugaz} {et~al.}(2015){Lugaz}, {Farrugia}, {Smith}, \&
  {Paulson}}]{Lugaz2015}
{Lugaz}, N., {Farrugia}, C.~J., {Smith}, C.~W., \& {Paulson}, K. 2015, Journal
  of Geophysical Research (Space Physics), 120, 2409

\bibitem[{{Lugaz} {et~al.}(2005){Lugaz}, {Manchester}, \&
  {Gombosi}}]{Lugaz2005}
{Lugaz}, N., {Manchester}, IV, W.~B., \& {Gombosi}, T.~I. 2005, \apj, 634, 651

\bibitem[{{Lugaz} {et~al.}(2009){Lugaz}, {Vourlidas}, \& {Roussev}}]{Lugaz2009}
{Lugaz}, N., {Vourlidas}, A., \& {Roussev}, I.~I. 2009, Annales Geophysicae,
  27, 3479

\bibitem[{{Mart{\'{\i}}nez Oliveros} {et~al.}(2012){Mart{\'{\i}}nez Oliveros},
  {Raftery}, {Bain}, {Liu}, {Krupar}, {Bale}, \&
  {Krucker}}]{Martinez-Oliveros2012}
{Mart{\'{\i}}nez Oliveros}, J.~C., {Raftery}, C.~L., {Bain}, H.~M., {et~al.}
  2012, \apj, 748, 66

\bibitem[{{Mishra} \& {Srivastava}(2014)}]{Mishra2014a}
{Mishra}, W., \& {Srivastava}, N. 2014, \apj, 794, 64

\bibitem[{{Mishra} {et~al.}(2015{\natexlab{a}}){Mishra}, {Srivastava}, \&
  {Chakrabarty}}]{Mishra2015}
{Mishra}, W., {Srivastava}, N., \& {Chakrabarty}, D. 2015{\natexlab{a}},
  \solphys, 290, 527

\bibitem[{{Mishra} {et~al.}(2014){Mishra}, {Srivastava}, \&
  {Davies}}]{Mishra2014}
{Mishra}, W., {Srivastava}, N., \& {Davies}, J.~A. 2014, \apj, 784, 135

\bibitem[{{Mishra} {et~al.}(2015{\natexlab{b}}){Mishra}, {Srivastava}, \&
  {Singh}}]{Mishra2015a}
{Mishra}, W., {Srivastava}, N., \& {Singh}, T. 2015{\natexlab{b}}, Journal of
  Geophysical Research (Space Physics), 120, 10

\bibitem[{{M{\"o}stl} {et~al.}(2012){M{\"o}stl}, {Farrugia}, {Kilpua}, {Jian},
  {Liu}, {Eastwood}, {Harrison}, {Webb}, {Temmer}, {Odstrcil}, {Davies},
  {Rollett}, {Luhmann}, {Nitta}, {Mulligan}, {Jensen}, {Forsyth}, {Lavraud},
  {de Koning}, {Veronig}, {Galvin}, {Zhang}, \& {Anderson}}]{Mostl2012}
{M{\"o}stl}, C., {Farrugia}, C.~J., {Kilpua}, E.~K.~J., {et~al.} 2012, \apj,
  758, 10

\bibitem[{{Newton}(1687)}]{Newton1687}
{Newton}, I. 1687, {Philosophiae Naturalis Principia Mathematica. Auctore Js.
  Newton}, doi:10.3931/e-rara-440

\bibitem[{{Niembro} {et~al.}(2015){Niembro}, {Cant{\'o}}, {Lara}, \&
  {Gonz{\'a}lez}}]{Niembro2015}
{Niembro}, T., {Cant{\'o}}, J., {Lara}, A., \& {Gonz{\'a}lez}, R.~F. 2015,
  \apj, 811, 69

\bibitem[{{Rouillard} {et~al.}(2008){Rouillard}, {Davies}, {Forsyth}, {Rees},
  {Davis}, {Harrison}, {Lockwood}, {Bewsher}, {Crothers}, {Eyles}, {Hapgood},
  \& {Perry}}]{Rouillard2008}
{Rouillard}, A.~P., {Davies}, J.~A., {Forsyth}, R.~J., {et~al.} 2008, \grl, 35,
  10110

\bibitem[{{Schmidt} \& {Cargill}(2004)}]{Schmidt2004}
{Schmidt}, J., \& {Cargill}, P. 2004, Annales Geophysicae, 22, 2245

\bibitem[{{Sheeley} {et~al.}(1999){Sheeley}, {Walters}, {Wang}, \&
  {Howard}}]{Sheeley1999}
{Sheeley}, N.~R., {Walters}, J.~H., {Wang}, Y.-M., \& {Howard}, R.~A. 1999,
  \jgr, 104, 24739

\bibitem[{{Shen} {et~al.}(2012){Shen}, {Wang}, {Wang}, {Liu}, {Liu},
  {Vourlidas}, {Miao}, {Ye}, {Liu}, \& {Zhou}}]{Shen2012}
{Shen}, C., {Wang}, Y., {Wang}, S., {et~al.} 2012, \nat, 8, 923

\bibitem[{{Shen} {et~al.}(2013){Shen}, {Shen}, {Wang}, {Feng}, \&
  {Xiang}}]{Shen2013}
{Shen}, F., {Shen}, C., {Wang}, Y., {Feng}, X., \& {Xiang}, C. 2013, \grl, 40,
  1457

\bibitem[{{Shen} {et~al.}(2014){Shen}, {Shen}, {Zhang}, {Hess}, {Wang}, {Feng},
  {Cheng}, \& {Yang}}]{Shen2014}
{Shen}, F., {Shen}, C., {Zhang}, J., {et~al.} 2014, Journal of Geophysical
  Research (Space Physics), 119, 7128

\bibitem[{{Shen} {et~al.}(2016){Shen}, {Wang}, {Shen}, \& {Feng}}]{Shen2016}
{Shen}, F., {Wang}, Y., {Shen}, C., \& {Feng}, X. 2016, Scientific Reports, 6,
  19576

\bibitem[{{Stewart}(2000)}]{Stewart2000}
{Stewart}, D.~E. 2000, SIAM Review, 42, 3

\bibitem[{{Temmer} \& {Nitta}(2015)}]{Temmer2015}
{Temmer}, M., \& {Nitta}, N.~V. 2015, \solphys, 290, 919

\bibitem[{{Temmer} {et~al.}(2012){Temmer}, {Vr{\v s}nak}, {Rollett}, {Bein},
  {de Koning}, {Liu}, {Bosman}, {Davies}, {M{\"o}stl}, {{\v Z}ic}, {Veronig},
  {Bothmer}, {Harrison}, {Nitta}, {Bisi}, {Flor}, {Eastwood}, {Odstrcil}, \&
  {Forsyth}}]{Temmer2012}
{Temmer}, M., {Vr{\v s}nak}, B., {Rollett}, T., {et~al.} 2012, \apj, 749, 57

\bibitem[{{Thernisien} {et~al.}(2009){Thernisien}, {Vourlidas}, \&
  {Howard}}]{Thernisien2009}
{Thernisien}, A., {Vourlidas}, A., \& {Howard}, R.~A. 2009, \solphys, 256, 111

\bibitem[{{Vandas} {et~al.}(1997){Vandas}, {Fischer}, {Dryer}, {Smith},
  {Detman}, \& {Geranios}}]{Vandas1997}
{Vandas}, M., {Fischer}, S., {Dryer}, M., {et~al.} 1997, \jgr, 102, 22295

\bibitem[{{Vandas} \& {Odstrcil}(2004)}]{Vandas2004}
{Vandas}, M., \& {Odstrcil}, D. 2004, \aap, 415, 755

\bibitem[{{Vemareddy} \& {Mishra}(2015)}]{Vemareddy2015}
{Vemareddy}, P., \& {Mishra}, W. 2015, \apj, 814, 59

\bibitem[{{Wang} {et~al.}(2009){Wang}, {Zhang}, \& {Shen}}]{Wang2009}
{Wang}, Y., {Zhang}, J., \& {Shen}, C. 2009, Journal of Geophysical Research
  (Space Physics), 114, A10104

\bibitem[{{Wang} {et~al.}(2005){Wang}, {Zheng}, {Wang}, \& {Ye}}]{Wang2005}
{Wang}, Y., {Zheng}, H., {Wang}, S., \& {Ye}, P. 2005, \aap, 434, 309

\bibitem[{{Wang} {et~al.}(2003){Wang}, {Ye}, {Wang}, \& {Xue}}]{Wang2003a}
{Wang}, Y.~M., {Ye}, P.~Z., {Wang}, S., \& {Xue}, X.~H. 2003, \grl, 30, 1700

\bibitem[{{Webb} {et~al.}(2013){Webb}, {M{\"o}stl}, {Jackson}, {Bisi},
  {Howard}, {Mulligan}, {Jensen}, {Jian}, {Davies}, {de Koning}, {Liu},
  {Temmer}, {Clover}, {Farrugia}, {Harrison}, {Nitta}, {Odstrcil}, {Tappin}, \&
  {Yu}}]{Webb2013}
{Webb}, D.~F., {M{\"o}stl}, C., {Jackson}, B.~V., {et~al.} 2013, \solphys, 285,
  317

\bibitem[{{Wood} {et~al.}(2010){Wood}, {Howard}, \& {Socker}}]{Wood2010}
{Wood}, B.~E., {Howard}, R.~A., \& {Socker}, D.~G. 2010, \apj, 715, 1524

\bibitem[{{Xiong} {et~al.}(2009){Xiong}, {Zheng}, \& {Wang}}]{Xiong2009}
{Xiong}, M., {Zheng}, H., \& {Wang}, S. 2009, Journal of Geophysical Research
  (Space Physics), 114, 11101

\bibitem[{{Xiong} {et~al.}(2006){Xiong}, {Zheng}, {Wang}, \&
  {Wang}}]{Xiong2006}
{Xiong}, M., {Zheng}, H., {Wang}, Y., \& {Wang}, S. 2006, Journal of
  Geophysical Research (Space Physics), 111, 11102

\bibitem[{{Xiong} {et~al.}(2007){Xiong}, {Zheng}, {Wu}, {Wang}, \&
  {Wang}}]{Xiong2007}
{Xiong}, M., {Zheng}, H., {Wu}, S.~T., {Wang}, Y., \& {Wang}, S. 2007, Journal
  of Geophysical Research (Space Physics), 112, 11103

\end{thebibliography}



\end{document}